\documentclass[english]{article}
\usepackage[T1]{fontenc}
\usepackage[latin9]{inputenc}
\usepackage{geometry}
\usepackage{xcolor}
\usepackage{pdfcolmk}
\usepackage{babel}
\usepackage{amsmath}
\usepackage{amssymb}
\usepackage{slashed}
\usepackage{graphicx}
\PassOptionsToPackage{normalem}{ulem}
\usepackage{ulem}
\usepackage[unicode=true,pdfusetitle,
 bookmarks=true,bookmarksnumbered=false,bookmarksopen=false,
 breaklinks=false,pdfborder={0 0 0},pdfborderstyle={},backref=false,colorlinks=false]
 {hyperref}

\makeatletter

\providecolor{lyxadded}{rgb}{0,0,1}
\providecolor{lyxdeleted}{rgb}{1,0,0}

\DeclareRobustCommand{\lyxsout}[1]{\ifx\\#1\else\sout{#1}\fi}

\newcommand{\lyxaddress}[1]{
	\par {\raggedright #1
	\vspace{1.4em}
	\noindent\par}
}

\usepackage{xcolor}
\hypersetup{
    colorlinks,
    linkcolor={red!50!black},
    citecolor={blue!50!black},
    urlcolor={blue!50!black}
}

\makeatother

\begin{document}
\title{L\"uscher-corrections for 1-particle form-factors in non-diagonally scattering integrable quantum field theories}
\author{\'Arp\'ad Heged\H{u}s}
\maketitle

\lyxaddress{\begin{center}
~\\
Wigner Research Centre for Physics\\
Konkoly-Thege Mikl\'os u. 29-33, 1121 Budapest, Hungary\\
\par\end{center}}
\begin{abstract}
In this paper we derive from field theory a L\"uscher-formula, which gives the leading exponentially small  
in volume corrections
 to the 1-particle form-factors in non-diagonally scattering integrable quantum field theories. 
Our final formula is expressed in terms of appropriate expressions of 1- and 3-particle form-factors, and can be considered as the generalization of previous results obtained for diagonally scattering bosonic integrable quantum field theories. Since our formulas are also valid for fermions and operators with non-zero Lorentz-spin, 
we demonstrated our results in the Massive Thirring Model, and checked our formula against 1-loop perturbation theory finding perfect agreement. 
\end{abstract}

\section{Introduction} \label{intro}

The understanding of finite size physics in Integrable Quantum Field Theory, 
became important in the AdS/CFT correspondence \cite{Beisert:2010jr} 
and in modern statistical and condensed matter physics applications \cite{KoEss}.

In the past decades, useful and efficient mathematical tools have been developed to describe the 
exact finite volume dependence of the spectrum of these theories. 
Concentrating only to the main achievements, these are the so-called L\"uscher-formulas \cite{Luscher1,Luscher2} 
for the leading \cite{BaJa} and next to leading \cite{Bombardelli:2013yka} 
order exponentially small in volume corrections, 
the Thermodynamic Bethe Ansatz \cite{Zamolodchikov:1989cf} for the ground state and for excited states 
\cite{Dorey:1996re,BHO34,GKVsu2}, NLIE technique for some specific families of models \cite{KP1,ddv92,SuzS1,KLsun,BHnlie},
 and finally the Quantum Spectral Curve method for the different variants of the planar AdS/CFT correspondence \cite{QSC54,QSC43}.

The next step towards the bootstrap solution of an 
integrable quantum field theory in finite volume, is the description of the finite volume dependence of the matrix elements of the operators of the theory. 
The knowledge of such finite volume form-factors becomes important in spectral representation 
of thermal correlators \cite{PT10T,Gohmann}, and in the determination of the 
string field theory vertex \cite{BJsftv} and of the 3-point functions  \cite{KomatsuLect} in AdS/CFT.

The approach initiated in \cite{PT08a,PT08b}, sought for finite volume matrix elements in a 
form of a large volume series built out of the infinite volume form-factors of the theory. 
In these works,  polynomial in the inverse of the volume corrections to the 
form-factors, coming from the Bethe-Yang quantization rules of the rapidities, have been determined.

In general, integrable quantum field theories can be divided into two classes; diagonally scattering (or purely elastic scattering) theories, 
and non-diagonally scattering theories. In the former, there is no mass degeneration in the spectrum, and the 
  S-matrix of any 2-particle scattering is a pure phase factor (c-number). In a non-diagonally scattering theory, 
in general there is some global symmetry present, which causes mass degeneration in the spectrum. 
As a consequence, particles carry flavor quantum numbers, and the S-matrix becomes a non-diagonal 
matrix in the flavor space. This matrix structure makes the solution of the finite volume problem much more 
complex, than it is in the purely elastic scattering case.

In purely elastic scattering theories a remarkable progress have been made in describing the finite volume 
behavior of form-factors. Namely, LeClair-Mussardo series have been conjectured
 for the diagonal matrix elements \cite{LM99,PST14} and for the 2-point functions \cite{PSZLM}. 
In the special case of the sinh-Gordon model, solving a set of linear integral equations 
allows one to determine, the exact finite volume expectation values \cite{SmirNeg,SmirBajn}, 
which can be considered as 
a more compact resummation of the LeClair-Mussardo series. 
As for the non-diagonal matrix elements, the leading order L\"uscher-corrections 
have been conjectured{\footnote{To be more precise, the so-called F-term has been conjectured in 
\cite{BBCL,BBCL1}, and the $\mu$-term in \cite{Pomu}.}} \cite{Pomu,BBCL,BBCL1}, 
but an exact expression resumming all exponentially small in volume corrections is still missing.

Unfortunately, much less is known in the case of non-diagonally scattering theories. 
The exact determination of diagonal matrix elements of local operators is known only in some very 
specific models like the sine-Gordon \cite{SGV,SGen} or $N=1$ super sine-Gordon theories \cite{SmirBab}. 
In the latter case, only the vacuum expectation values have been worked out. 
Similarly, the LeClair-Mussardo series representation for 2-point functions is known only 
for some non-relativistic model with $gl(3)$ or $gl(2|1)$ symmetry \cite{HuPoPri}.

The finite volume corrections to the non-diagonal form-factors in this class of 
theories are much less known. So far, nothing is known about the corrections 
to their Bethe-Yang limit \cite{PT08a}. To fill this gap, and reach some progress in this direction, 
in this paper we aim to derive the leading order exponentially small in volume corrections 
(L\"uscher-corrections) to the 1-particle form-factors in a general relativistically invariant 
integrable quantum field theory. We achieve this plan, by an appropriate generalization of the field theoretical method 
initiated in \cite{BBCL} for purely elastic scattering theories.

The paper is organized as follows:
In section \ref{sect2}, the method of extracting the L\"uscher-corrections to finite-volume 
form-factors is described. In section \ref{sect3}, we recall the form-factor axioms being 
necessary to the computations. In section \ref{sect4}, we compute the L\"uscher-correction to the 
1-particle form-factors using the framework of the mirror theory. For the perturbative check 
of our final result, we summarize some important properties of the massive Thirring model in section 
\ref{sect5}. 
In section \ref{sect6}, the leading order weak coupling expression of the L\"uscher-correction to the 
fermion propagator of the Massive Thirring model at the 1-particle pole is determined from the weak coupling 
expansion of our exact results. The same quantity is computed  at 1-loop order from Lagrangian perturbation theory 
in section \ref{sect7}.  
The paper is closed by the summary of results in section \ref{summary}. 
Some list of formulas are relegated to appendix \ref{appA}.

\section{The 2-point function and finite volume form-factors} \label{sect2}

In this section guided by the method of \cite{BBCL}, we recall how one can extract 
the finite volume corrections of the form-factors from the 2-point functions of the corresponding operators. 
This requires to compute the 2-point function in two different ways. First, directly in the "finite-volume channel", 
and then in the so-called mirror-channel, where the role of space and time is interchanged with respect to 
the previous channel. 
We consider, the finite volume 2-point function as follows: 
\begin{equation} \label{2ptv0}
\begin{split}
\langle \bar{\psi}_\alpha(x,t) \psi_{\beta}(0) \rangle_L=
\frac{\int[{\cal D}\psi] \, \bar{\psi}_\alpha(x,t) \psi_{\beta}(0)\, e^{-S[\psi]}}{\int[{\cal D}\psi] e^{-S[\psi]}},
\end{split}
\end{equation}
where $S[\psi]$ is the Euclidean action, and in the path integral representation
 the field configurations are either periodic or anti-periodic 
in $x$ with $L,$ depending on whether the fields are commuting or anti-commuting ones. In our case the time 
coordinate $t$ is not compactified and can take any real value. For our purposes we need the 2-point function in 
momentum space:
\begin{equation} \label{Gmom}
\begin{split}
\Gamma_{\alpha \beta}(\omega,q)=\frac{1}{L} \, \int\limits_{-L/2}^{L/2}dx \, 
\int\limits_{-\infty}^{\infty} dt \, e^{i\, \omega t+i \, q x} \, \langle \bar{\psi}_\alpha(x,t) \psi_{\beta}(0) \rangle_L.
\end{split}
\end{equation}
The periodicity of the fields imposes the constraint on the momentum: $e^{i \, q L}=(-1)^F,$ 
where $F=0$ for bosonic fields and $F=1$ for fermionic fields. 
For the sake of simplicity, in this paper we will consider models having only either fermions or 
bosons in their spectrum, and assume that the fields entering the 2-point function (\ref{2ptv0})
share the same statistics with the fundamental fields of the Lagrangian of the theory.
In the "finite-volume channel", where the periodic $x$ is interpreted as a space coordinate, the 
2-point function is a time ordered product:
\begin{equation} \label{Tprod}
\begin{split}
\langle \bar{\psi}_\alpha(x,t) \psi_{\beta}(0) \rangle_L=\theta(t)\, \langle \bar{\psi}_\alpha(x,t) \psi_{\beta}(0) \rangle_L+(-1)^F\, \theta(-t) \langle \psi_{\beta}(0) \bar{\psi}_\alpha(x,t)  \rangle_L,
\end{split}
\end{equation}
with $\theta(t)$ being the Heaviside-function.
After the insertion a complete set of finite volume eigenstates, this form implies the following 
representation for $\Gamma_{\alpha \beta}(\omega,q):$ 
\begin{equation} \label{GM1}
\begin{split}
\Gamma_{\alpha \beta}(\omega,q)&=\sum\limits_{\{n,A\}} \, 
{}_L\langle 0|{\cal P} \bar{\psi}_\alpha(0) {\cal P}^{-1}|n\rangle_{A}\,\,
{}_A\langle n|{\cal P} {\psi}_\beta(0) {\cal P}^{-1}|0\rangle_L\, 
\frac{i \, \delta_{q,P_n(L)}}{\omega+i \, \Delta E_n(L)}+\\
&+\sum\limits_{\{n,A\}} \, 
{}_L\langle 0|{\psi}_\beta(0)|n\rangle_{A}\,\,
{}_A\langle n| \bar{\psi}_\alpha(0) |0\rangle_L\, 
\frac{-i (-1)^F \, \delta_{q,P_n(L)}}{\omega-i \, \Delta E_n(L)},
\end{split}
\end{equation}
where ${\cal P}$ denotes the parity transformation, $\Delta E_n(L)$ and $P_n(L)$ 
are the energy and momentum of the state $|n\rangle_A,$ such that the energy is measured from the 
finite volume ground state energy: $\Delta E_n(L)=E_n(L)-E_0(L).$ 
Here, the subscript $A$ accounts for the degeneracies dictated by the global symmetries of the model.
This shows, that the finite-volume form-factors can be read off from the residues of the poles of 
$\Gamma_{\alpha \beta}(\omega, q)$ in the variable $\omega.$ 
Now, we specify this statement to the 1-particle pole at $\omega=\Delta E_1(L)$: 
 \begin{equation} \label{1partGM}
\begin{split}
\underset{\omega=i \Delta E_1(L)}{\mbox{Res}}\, \Gamma_{\alpha \beta}(\omega,q)=-i \, (-1)^F \, \sum\limits_a {}_L\langle 0|{\psi}_\beta(0)|q \rangle_{a}\,\,
{}_a\langle q| \bar{\psi}_\alpha(0) |0\rangle_L,  
\end{split}
\end{equation}
where the 1-particle states are indexed by their momentum $q$ and their flavor index $a.$
Formula (\ref{1partGM}) contains a sum on the right hand side, which doesn't seem to allow one to 
extract the individual form-factors. Nevertheless, in most of the interesting cases 
as a consequence of the global symmetry of the model, the 1-particle  
form-factors have a Kronecker-delta structure in the flavor space. 
Namely, ${}_L\langle 0|{\psi}_\beta(0)|q \rangle_{a}\sim \delta_{a, x_0},$ 
with $x_0$ being a specific particle of the lowest lying multiplet of the theory. 
Specific examples for this case are the fundamental fields of the $O(N)$ nonlinear $\sigma$-models or 
of the massive Thirring model. 
Then (\ref{1partGM}) takes a simpler form, which allows one to determine products of form-factors:
\begin{equation} \label{1partGMv1}
\begin{split}
\underset{\omega=i \Delta E_1(L)}{\mbox{Res}}\, \Gamma_{\alpha \beta}(\omega,q)=-i \, (-1)^F \,  {}_L\langle 0|{\psi}_\beta(0)|q \rangle_{x_0}\,\,
{}_{x_0}\langle q| \bar{\psi}_\alpha(0) |0\rangle_L.  
\end{split}
\end{equation}
The situation is even better, when the exponentially small in volume corrections are sought for. 
Denote the Bethe-Yang part of a form-factor by:  $\langle 0|{\psi}_\beta(0)|q \rangle_{a}^{BY}$ and the 
L\"uscher-correction by $\delta\langle 0|{\psi}_\beta(0)|q \rangle_{a}.$ 
Then the L\"uscher order of the residue in (\ref{1partGMv1}) can be written as: 
\begin{equation} \label{1partGMlusi}
\begin{split}
\underset{\omega=i \Delta E_1(L)}{\mbox{Res}}\, \delta\Gamma_{\alpha \beta}(\omega,q)=-i \, (-1)^F \, \left\{ 
\langle 0|{\psi}_\beta(0)|q \rangle_{x_0}^{BY} \, \delta ({}_{x_0}\langle q| \bar{\psi}_\alpha(0) |0\rangle)
+\right.\\ 
\left.{}_{x_0}\langle q| \bar{\psi}_\alpha(0) |0\rangle^{BY} \delta\langle 0|{\psi}_\beta(0)|q \rangle_{x_0} \, 
\right\}.
\end{split}
\end{equation} 
Since the correction of one form-factor is multiplied with the Bethe-Yang limit of the other, 
at the formal level this formula allows one to extract each 1-particle form-factor's  
exponential correction, from the L\"uscher-order expression of the residue of the 2-point function.

To express the L\"uscher corrections of the form-factors in terms of infinite volume data, the same 2-point 
function and residue must be calculated in the language of the mirror-model \cite{BBCL}. 
Since, the mirror model is obtained from the original one by exchanging the role of space and time, 
 in this language the original 2-point function becomes a thermal correlator with $T=1/L$ temperature. 
In general the fields transform under the 
mirror transformation, thus in the mirror framework both the action and the fields defining the 2-point function 
may differ from the original ones{\footnote{In a section \ref{sect5}, this statement may become clearer for the reader, when the example of the massive Thirring model will be discussed.}}.
Denote the mirror transforms of the fields entering (\ref{2ptv0}):
\begin{equation} \label{mirtf}
\begin{split}
\psi_\beta &\rightarrow {\cal M}(\psi_\beta)=\phi_\beta, \\
\bar{\psi}_\alpha &\rightarrow {\cal M}(\bar{\psi}_\alpha)=\bar{\phi}_\alpha.
\end{split}
\end{equation}
Then, the original correlator can be computed in the mirror channel by the usual thermal formula:
\begin{equation} \label{thermal}
\begin{split}
\langle \bar{\psi}_\alpha(x,t) \psi_{\beta}(0) \rangle_L =
\langle \bar{\phi}_\alpha(x,t) \phi_{\beta}(0) \rangle_{T=1/L}&=\\
\frac{\theta(x)}{Z}\, \mbox{Tr}\left[\bar{\phi}_{\alpha}(0,t) \, e^{-\bar{H} x} \, \phi_\beta(0) \, e^{-\bar{H}(L-x)} \right]&+
\frac{(-1)^F\,\theta(-x)}{Z}\, \mbox{Tr}\left[\bar{\phi}_{\beta}(0) \, e^{\bar{H} x} \, \bar{\phi}_{\alpha}(0,t) \, e^{-\bar{H}(L+x)} \right],
\end{split}
\end{equation}
where $\bar{H}$ denotes the mirror Hamiltonian and $Z=\mbox{Tr}\left[ e^{-\bar{H}L}\right]$ is the mirror partition function. 
Inserting a complete set of mirror eigenstates, one ends up with the formula:
\begin{equation} \label{ZGmir}
\begin{split}
\Gamma_{\alpha \beta}(\omega,q)=\frac{1}{Z} \left\{\Sigma(\bar{\phi}_\alpha,\phi_\beta|q)+(-1)^F \, \Sigma(\phi_\beta^{\cal P},\bar{\phi}_\alpha^{\cal P}|-q) \right\},
\end{split}
\end{equation}
where for any operator $\phi,$ $\phi^{\cal P}$ stands for its parity transform: ${\cal P} \phi {\cal P}^{-1},$ and 
\begin{equation} \label{Sigma}
\begin{split}
\Sigma(\bar{\phi}_\alpha,\phi_\beta|q)=\frac{2 \pi}{L} \, \sum\limits_n e^{-E_n L} \sum\limits_m \,
T_{nm}[\bar{\phi}_\alpha,\phi_\beta] \, \frac{\delta(\omega+P_{mn})}{E_{mn}-i q},
\end{split}
\end{equation}
where $E_{mn}=E_n-E_m, P_{mn}=P_n-P_m,$ and  
\begin{equation} \label{Tnm}
\begin{split}
T_{nm}[\bar{\phi}_\alpha,\phi_\beta]=\langle n|\bar{\phi}_\alpha |m \rangle \langle m|{\phi}_\beta |n \rangle.
\end{split}
\end{equation}
Here the states $|n\rangle$ and $|m\rangle$ are eigenstates of the infinite volume mirror Hamiltonian $\bar{H}.$
As it can be seem from (\ref{ZGmir}) and (\ref{Sigma}), 
the mirror representation is particularly useful for computing the exponentially 
small in volume corrections to the 2-point function, since they are governed by the $\sum\limits_n e^{-E_n L}...$ factor in (\ref{Sigma}). Thus for leading L\"uscher-corrections the sum for $n$ should be restricted to the 
1-particle states with smallest mass of the mirror theory. 
In the next sections, from (\ref{ZGmir}), we express the L\"uscher-correction 
of the residue at the 1-particle pole in terms of infinite volume data. 
To do so, first we have to fix our conventions on the S-matrix and form-factors. 

\section{Form-factor axioms and related conventions} \label{sect3}

In an integrable relativistic quantum field theory, the 2-particle S-matrix plays crucial role.
Denote $a(\theta)$ a particle with flavor $a$ and rapidity $\theta.$ Then, in our notation the scattering process:
$a(\theta_1)+b(\theta_2)\rightarrow c(\theta_1)+d(\theta_2)$ is described by the S-matrix element: 
${ S}^{cd}_{ab}(\theta_1-\theta_2).$ 
In the sequel we assume, that the S-matrix satisfies the following properties:
\begin{eqnarray} & \bullet & \text{Parity-symmetry:} \qquad  \qquad \qquad
{ S}_{ab}^{cd}(\theta)={ S}_{ba}^{dc}(\theta), \label{Bose} \\
 & \bullet & \text{Time-reversal symmetry:} \qquad  \quad \!
{ S}_{ab}^{cd}(\theta)={ S}_{cd}^{ab}(\theta), \label{Time} \\
 & \bullet & \text{Crossing-symmetry:} \qquad  \qquad \quad
{ S}_{ab}^{cd}(\theta)={ S}_{a\bar{d}}^{c\bar{b}}(i \, \pi-\theta), \label{Cross} \\
 & \bullet & \text{Unitarity:} \qquad \qquad \qquad \, \,
{ S}_{ab}^{ef}(\theta)\, { S}_{ef}^{cd}(-\theta)=\delta_a^c \, \delta_b^d, \label{Unit} \\
 & \bullet & \text{Real analyticity:} \qquad  \qquad \quad \quad \!
{ S}_{ab}^{cd}(\theta)^*={ S}_{ab}^{cd}(-\theta^*),  \label{Real} 
\end{eqnarray}
where $*$ denotes complex conjugation and for any index $a,$ $\bar{a}$ denotes the charge conjugated particle. 

In order to formulate the form factor axioms, we need to define the charge conjugation matrix, 
which gives how the charge conjugation acts on 1-particle states:
\begin{equation} \label{Ckonj}
\begin{split}
C_{ab}=\delta_{a\bar{b}}=\delta_{\bar{a}b}.
\end{split}
\end{equation}
We use the following normalization for the infinite volume multi-particle states:
\begin{equation} \label{StateNorm}
\begin{split}
{}_{b_1...b_n}\langle \theta'_1,...,\theta'_n|\theta_1,...,\theta_n \rangle_{a_1,...,a_n}=
\prod\limits_{j=1}^n \delta_{a_j b_j} \, \delta(\theta'_j-\theta_j).
\end{split}
\end{equation}
We would like to formulate the form-factor axioms to be valid for fermions, too.  
For this reason, following \cite{BABK1}, it is worth introducing the dotted S-matrix with the definition as follows: 
\begin{equation} \label{Sdot}
\begin{split}
\dot{S}_{ab}^{cd}(\theta)=(-1)^F S_{ab}^{cd}(\theta),
\end{split}
\end{equation}
where we recall that $F=0, \mbox{and} \,\, 1$ for bosonic and fermionic theories, 
respectively{\footnote{We note, that for the sake of simplicity in this paper,
 we consider theories with all particles being either bosons or fermions.}}.  

With these conventions the form-factors $F^{{\cal O}}$ of a local operator ${\cal O}(x,t)$ 
which is local with respect to all fields creating the particles of the model,    
satisfy the axioms as follows \cite{Smirnov}:
\newline{I. Lorentz-invariance:}
\begin{equation} \label{ax1}
\begin{split}
F^{\cal O}_{a_1...a_n}(\theta_1+\theta,...,\theta_n+\theta)=e^{s_{\cal O} \theta} \, F^{\cal O}_{a_1...a_n}(\theta_1,...,\theta_n),
\end{split}
\end{equation}
where $s_{\cal O}$ is the Lorentz-spin of ${\cal O}.$
\newline{II. Exchange:}
\begin{equation} \label{ax2}
\begin{split}
F^{\cal O}_{...a_j a_{j+1}...}(...,\theta_j,\theta_{j+1},...)={\dot{S}}_{a_j a_{j+1}}^{b_{j+1} b_j}(\theta_j-\theta_{j+1}) \, F^{\cal O}_{...b_{j} b_{j+1}...}(...,\theta_{j+1},\theta_{j},...),
\end{split}
\end{equation}
\newline{III. Cyclic permutation:}
\begin{equation} \label{ax3}
\begin{split}
F^{\cal O}_{a_1 a_2...a_n}(\theta_1+2 \pi \, i,...,\theta_n)=\sigma_{{\cal O},1}\, 
F^{\cal O}_{a_2...a_n a_1}(\theta_2,...,\theta_n,\theta_1),
\end{split}
\end{equation}
where $\sigma_{{\cal O},1}=-1,$ if the field creating the $a_1$ particle anticommutes with ${\cal O}$, and 
$\sigma_{{\cal O},1}=1,$ otherwise.
\newline{IV. Kinematical singularity:}
\begin{equation} \label{ax4}
\begin{split}
\underset{\theta'=\theta}{\mbox{Res}} \, F^{\cal O}_{ab u_1...u_n}(\theta'+i\, \pi,\theta, \theta_1,...,\theta_n)=
\frac{i}{2 \pi} \bigg\{ 
C_{ab} \, F^{\cal O}_{u_1...u_n}( \theta_1,...,\theta_n)-\\ 
\sum\limits_{v_1,..,v_n}{\cal T}_{b}^{\bar{a}}(\theta|\theta_1,..,\theta_n)_{u_1...u_n}^{v_1...v_n} \,
F^{\cal O}_{v_1...v_n}(\theta_1,...,\theta_n)
\bigg\},
\end{split}
\end{equation}
where ${\cal T}$ denotes the soliton monodromy matrix defined by:
\begin{equation} \label{monodr}
\begin{split}
{\cal T}_a^b(\theta|\theta_1,...,\theta_n)_{a_1 a_2 ... a_n}^{b_1 b_2 ...b_n}=
{ S}_{a \, a_1}^{k_1 \, b_1}(\theta-\theta_1)
{ S}_{k_1 \,  a_2}^{k_2 \, b_2}(\theta-\theta_2)...
{ S}_{k_{n-1}\,  a_n}^{b \, b_n}(\theta-\theta_n).
\end{split}
\end{equation}
In this paper we focus on models with having no bound states of fundamental particles in their spectrum, 
thus we don't need to recall the dynamical-pole axiom. What is needed is the crossing axiom for form factors. 
With our conventions, it takes the form:
\begin{equation} \label{crosff}
\begin{split}
{}_{a_1}\langle \theta_1|{\cal O}|\theta_2,...,\theta_n & \rangle_{a_2,...,a_n}=\sigma_{{\cal O},1} \,
{\Big \{} F^{\cal O}_{\bar{a}_1,a_2,...,a_n}(\theta_1+i \, \pi+i \,\epsilon,\theta_2,...,\theta_n) + \\
+\sum\limits_{j=1}^n \delta(\theta_1-\theta_j) & \,  
 F^{\cal O}_{a'_2,...,a'_{j-1},a_{j+1},...a_n}(\theta_1,...,\theta_{j-1},\theta_{j+1},...,\theta_n) \, 
\dot{{\cal T}}_{a_j}^{a_1}(\theta_2,...,\theta_{j-1}|\theta_j)_{a_2 ... a_{j-1}}^{a'_2 ...a'_{j-1}} 
 {\Big \}}, 
\end{split}
\end{equation}
where $\epsilon$ is a positive infinitesimal number and
\begin{equation} \label{Tjdot}
\begin{split}
\dot{{\cal T}}_{a}^{b}(\theta_2,...,\theta_{j-1}|\theta_j)_{a_2 ... a_{j-1}}^{a'_2 ...a'_{j-1}}=
\dot{S}_{a_2 \, a}^{a'_2 \, b_2}(\theta_2-\theta_j)\, \dot{S}_{a_3 \, b_2}^{a'_3 \, b_3}(\theta_3-\theta_j)
...\, \dot{S}_{a_{j-1} \, b_{j-2}}^{a'_{j-1} \, b}(\theta_{j-1}-\theta_j).
\end{split}
\end{equation}

\section{L\"uscher-corrections from the mirror representation} \label{sect4}

As we mentioned in the previous sections, the mirror representation of the 2-point function 
is well suited for computing the exponentially small in volume corrections to the leading order
Bethe-Yang limit. 

In this section, we write down, how to extract the L\"uscher-corrections of the 1-particle 
form-factors and of the 1-particle energies. For short, we introduce the notations as follows:
\newline{$\bullet$ \emph{1-particle matrix elements of the original theory{\footnote{Here and in the sequel,  
 the term "original theory" means the theory where the compactified coordinate is interpreted as space, and the finite volume corrections in which we are interested in.}}
 in the Bethe-Yang limit:}}
\begin{equation} \label{fBYjel}
\begin{split}
f^\psi_a(\theta)=\langle 0| \psi| \theta\rangle_a^{BY}, \qquad 
\bar{f}^{\bar{\psi}}_a(\theta)={}_a \langle \theta| \bar{\psi}| 0\rangle^{BY}, 
\end{split}
\end{equation}
\newline{$\bullet$ \emph{1-particle matrix elements of the original theory at infinite volume:}}
\begin{equation} \label{fvegtelen}
\begin{split}
F^\psi_a(\theta)=\langle 0| \psi| \theta\rangle_a, \qquad 
\bar{F}^{\bar{\psi}}_a(\theta)={}_a \langle \theta| \bar{\psi}| 0\rangle,
\end{split}
\end{equation}
\newline{$\bullet$ \emph{Infinite volume matrix-elements in the mirror theory:}}
\begin{equation} \label{fFP}
\begin{split}
F_a^{\bar{\phi}}(\theta)=F_a(\theta), \qquad F_a^{\bar{\phi}^{\cal P}}(\theta)=F_a^P(\theta), \qquad 
F_a^{{\phi}}(\theta)=f_a(\theta), \qquad F_a^{{\phi}^{\cal P}}(\theta)=f_a^P(\theta),
\end{split}
\end{equation}
and similarly for any multiparticle form-factor in the mirror theory, for short we will simple use the notations:
 \begin{equation} \label{fFPmult}
\begin{split}
F^{\bar{\phi}}\to F, \qquad F^{\bar{\phi}^{\cal P}} \to F^P, \qquad 
F^{{\phi}} \to f, \qquad F^{{\phi}^{\cal P}} \to f^P.
\end{split}
\end{equation}
\newline{$\bullet$ \emph{Spin related phase factors:}}
\begin{equation} \label{sf}
\begin{split}
s_0=(-1)^F, \qquad s_f=e^{i\, \pi \, s_\phi},\qquad s_{f^P}=e^{i\, \pi \, s_{{\phi}^{\cal P}}},
\qquad s_{F}=e^{i\, \pi \, s_{\bar{\phi}}}
 \qquad s_{F^P}=e^{i\, \pi \, s_{\bar{\phi}^{\cal P}}},
\end{split}
\end{equation}
where $s_\phi,s_{{\phi}^{\cal P}}$ and $s_{\bar{\phi}},s_{\bar{\phi}^{\cal P}}$ 
are spins of the operators $\phi, \, \phi^{\cal P}$ and $\bar{\phi}, \,\bar{\phi}^{\cal P},$ 
respectively. We just recall that $\phi$ and $\bar{\phi}$ are the mirror counterparts of the original 
operators $\psi$ and $\bar{\psi},$ (cf.(\ref{mirtf})). For short, here and in the sequel, we skip to denote the 
subscript on the operators.

For later convenience, here we recall, that the relation between the Bethe-Yang (\ref{fBYjel}) and 
infinite volume (\ref{fvegtelen}) limits of form-factors is known from \cite{PT08a}. 
 The Bethe-Yang limit of a form-factor can be obtained as the ratio of its infinite volume counterpart 
and the square-root of the density of states corresponding to the sandwiching "ket" Bethe-Yang state. 
For 1-particle form-factors this means, that:
\begin{equation} \label{BYinf}
\begin{split}
f^\psi_a(\theta)=\langle 0|\psi|\theta\rangle_a^{BY}=\frac{F^\psi_a(\theta)}{\sqrt{\rho_1(\theta)}},
\end{split}
\end{equation}
with $\rho_1(\theta)$ being the appropriately normalized 1-particle density of Bethe-Yang states:
\begin{equation} \label{rho1}
\begin{split}
\rho_1(\theta)=\frac{m \, L}{2 \pi}\cosh \theta.
\end{split}
\end{equation}

Denote the L\"uscher-corrections of the 1-particle form-factors $f_a^{\psi}, \, \bar{f}^{\bar{\psi}}_a$ 
and of the 1-particle energy ${\cal E}(q)=\sqrt{q^2+m^2},$ by $\delta f_a^{\psi}, \, \delta\bar{f}^{\bar{\psi}}_a$ 
and $\delta{\cal E}(q),$ respectively. Then, from the large volume expansion of (\ref{GM1}) around the 1-particle 
pole, one obtains the following expression for the L\"uscher-correction:  
\begin{equation} \label{Gkifejt1}
\begin{split}
\Gamma^{(L)}(\omega,q)\sim\frac{{\cal Q}(\theta_{BY})}{\omega-i {\cal E}(q)}+
\frac{{\cal R}(\theta_{BY})}{\left(\omega-i {\cal E}(q)\right)^2},
\end{split}
\end{equation}
with
\begin{equation} \label{QR}
\begin{split}
{\cal Q}(\theta)&=-i \, (-1)^F \, \sum\limits_{a} \left[ f_a^{\psi}(\theta) \, \delta \bar{f}^{\bar{\psi}}_a(\theta)+
\bar{f}^{\bar{\psi}}_a(\theta) \,\delta f_a^{\psi}(\theta)\right], \\
{\cal R}(\theta)&= (-1)^F \,\delta {\cal E}(q) \,\sum\limits_{a} f_a^{\psi}(\theta) 
\bar{f}^{\bar{\psi}}_a(\theta), \\
\end{split}
\end{equation}
where $\theta_{BY}$ denotes the Bethe-Yang limit of the rapidity parameter. It is related to the momentum $q$ 
by
\begin{equation} \label{thBY}
\begin{split}
\theta_{BY}=\mbox{arcsinh}\left(\tfrac{q}{m}\right).
\end{split}
\end{equation}
In the cases, we are interested in, the 1-particle form-factors have a simple vector structure:
$f_a^{\psi}, \, \bar{f}^{\bar{\psi}}_a\sim \delta_{a,x_0},$ with $x_0$ being some specific 
flavor quantum number of the lowest lying multiplet of the theory. In this case the sums in 
(\ref{QR}) disappear and the expansion (\ref{Gkifejt1}) allows one to determine the L\"uscher-corrections of 
the 1-particle form-factors and the energy. Now, we perform this expansion in the mirror representation and 
express the L\"uscher-correction in terms of infinite volume scattering and form-factor data. 

As a 1st step, one should identify, which terms of the double sum (\ref{Sigma}) in (\ref{ZGmir}) 
contribute to the 1-particle pole of the 2-point function. In \cite{BBCL}, it has been shown, that 
\newline{{\bf in the Bethe-Yang limit:} $|n\rangle=|0\rangle$ and $|m\rangle$ runs for all 1-particle states, }
while {\bf in the 1st L\"uscher-order} two-types of contributions are possible:
\newline{{\bf 1.} $\quad |n\rangle$ runs for 1-particle states, and $|m\rangle=|0\rangle,$ }
\newline{{\bf 2.} $\quad |n\rangle$ runs for 1-particle states, and $|m\rangle$ runs for 2-particle states.}
These two types of contributions will be denoted by subscripts $10$ and $12,$ respectively.
\newline{Introducing the parameterizations{\footnote{Here $\psi$ denotes the rapidity parameter for $\omega,$ 
it shouldn't be confused with the field $\psi$ of the 2-point function (\ref{2ptv0}).}}:}
\begin{equation} \label{param}
\begin{split}
\omega=-m \sinh \psi \qquad \qquad q=m \, \hat{q},
\end{split}
\end{equation}
after a simple computation, the Bethe-Yang limit takes the form: 
\begin{equation} \label{GBY}
\begin{split}
\Gamma_{BY}(\omega,q)=\frac{2 \pi}{m^2 \, L} 
\left[ \frac{s_{F^P} \sum\limits_x f_x^P(\psi) \,F_{\bar{x}}^P(\psi) }{\cosh \psi\, (\cosh \psi+i \, \hat{q})} +
\frac{s_0 s_{f} \sum\limits_x F_x(\psi) \,f_{\bar{x}}(\psi) }{\cosh \psi\, (\cosh \psi-i \, \hat{q})}
\right].
\end{split}
\end{equation}
The $10$-type correction can also be easily computed. It takes the form:
\begin{equation} \label{GM10}
\begin{split}
\Gamma_{10}(\omega,q)=\Sigma_{10}(F,f|q)+s_0 \, \Sigma_{10}(f^{P},F^{P}|-q),
\end{split}
\end{equation}
where
\begin{equation} \label{S10}
\begin{split}
\Sigma_{10}(F,f|q)=-\frac{2 \pi}{m^2 L} \int\limits du \, e^{-\ell \cosh u} 
\frac{\delta(u-\psi)}{\cosh \psi \, (\cosh \psi +i \, \hat{q})}\, 
\sum\limits_a s_0 \, s_{F^P}\,f^P_{\bar{a}}(\psi) \, F^P_a(\psi),  
\end{split}
\end{equation}
where $\ell=m \, L$ and with this notation $\Sigma_{10}$ is considered as a functional of the form factors
 (\ref{fFP}). We note, that when taking into account the $(-1)$ sign factors, coming from commutation relations of 
the operators, we assumed, that all operators either commute or anticommute.
The most complicated $12$ term takes the form:
\begin{equation} \label{GM12}
\begin{split}
\Gamma_{12}(\omega,q)=\Sigma_{12}(F,f|q)+s_0 \, \Sigma_{12}(f^{P},F^{P}|-q),
\end{split}
\end{equation}
where $\Sigma_{12}$ can be formally written:
\begin{equation} \label{S12}
\begin{split}
\Sigma_{12}(F,f|q)=\frac{2 \pi}{m^2 \, L} \int\limits du \, e^{-\ell \, \cosh u} \, J_{12}(u,\psi|q)[F,f],
\end{split}
\end{equation}
with 
\begin{equation} \label{J12}
\begin{split}
J_{12}(u,\psi|q)[F,f]=
\int\limits_{-\infty}^{\infty} \!  \!d\beta_1 \!\! \int\limits_{-\infty}^{\infty} \! \! d\beta_2 
\, \frac{T(u|\beta_1,\beta_2)[F,f]}{2} \,
\frac{\sinh \beta_1+\sinh \beta_2-\sinh u -\sinh \psi}{\cosh \beta_1+\cosh \beta_2-\cosh u-i \hat{q}},
\end{split}
\end{equation}
where 
\begin{equation} \label{Tubb}
\begin{split}
T(u|\beta_,\beta_2)[F,f]=\sum\limits_{a,b_1,b_2} \Pi_{a b_1 b_2}(u|\beta_1,\beta_2)[F]\, \bar{\Pi}_{ b_2 b_1 a}(\beta_2,\beta_1|u)[f], 
\end{split}
\end{equation}
with the short notations for the necessary $12$ matrix elements of the operators:
\begin{equation} \label{pipibar}
\begin{split}
\Pi_{a b_1 b_2}(u|\beta_1,\beta_2)[F]&={}_a\langle u| \bar{\phi}|\beta_1,\beta_2 \rangle_{b_1 b_2}\\
\bar{\Pi}_{ b_2 b_1 a}(\beta_2,\beta_1|u)[f]&={}_{b_1 b_2}\langle \beta_1,\beta_2| {\phi}|u \rangle_{a}.
\end{split}
\end{equation}
As for the notation, we consider these matrix elements as functions of the form-factors of the 
corresponding operators.
These matrix elements can be computed from the form-factor axioms, and they take the form in the continuum theory
 as follows:
\begin{equation} \label{m1}
\begin{split}
\Pi_{a b_1 b_2}(u|\beta_1,\beta_2)[F]&=(-1)^F \, {\Big \{ }
F_{\bar{a} b_1 b_2}(u+i\, \pi^-,\beta_1,\beta_2)+\delta_{a b_1} f_{b_2}(\beta_2)+ \\
 &+\delta(u-\beta_2)\, \dot{S}_{b_1 b_2}^{x a}(\beta_1-\beta_2) \, F_{x}(\beta_2) {\Big \} }.
\end{split}
\end{equation}
\begin{equation} \label{m2}
\begin{split}
\bar{\Pi}_{ b_2 b_1 a}(\beta_2,\beta_1|u)[f]=
f_{\bar{b}_2 \bar{b}_1 a} (\beta_2+i \, \pi^-,\beta_1+i \, \pi^-,u)+
\delta(u-\beta_1) \, \delta_{a b_1} f_{\bar{b}_2}(\beta_2+i \, \pi)+\\
\delta(u-\beta_2) \, \dot{S}_{b_2 b_1}^{a x}(\beta_2-\beta_1) \, f_{\bar{x}}(\beta_1+i \, \pi),
\end{split}
\end{equation}
where for short $\pi^-=\pi-i \epsilon,$ with $\epsilon$ being a positive infinitesimal number, and for the  
repeated index $x$ summation is meant.

In addition to the so far discussed terms, the denominator $Z$ of (\ref{ZGmir}) also contributes to the 
L\"uscher-correction:
\begin{equation} \label{Zlus}
\begin{split}
Z=1+\Gamma_z+O(e^{-2\ell}), \qquad \Gamma_z=n_0 \, \delta(0) \, \int\limits du \, e^{-\ell \cosh u},
\end{split}
\end{equation}
where $n_0$ denotes the flavor-space dimension of the lowest lying multiplet of the theory.
Upto the L\"uscher order $O(e^{-\ell}),$ in the formulas (\ref{S10}),(\ref{S12}), and (\ref{Zlus}), 
the range of integration in $u$ can be restricted to the regime:
$\cosh u<2.$
This restriction becomes also very useful during the subsequent computations, 
since it allows us to avoid treating branch cuts. Thus, in the rest of the paper, everywhere 
where integration with respect to the variable $u$ can be seen, the following integration domain 
is meant: 
\begin{equation} \label{inturange}
-\mbox{arccosh} \, 2 <u<\mbox{arccosh} \, 2.
\end{equation}
Putting all the building blocks together,
 the L\"uscher correction to the 2-point function can be given as follows: 
\begin{equation} \label{GMlus}
\begin{split}
\Gamma^{(L)}(\omega,q)=-\Gamma_z\, \Gamma_{BY}(\omega,q)+\Gamma_{10}(\omega,q)+\Gamma_{12}(\omega,q)+
\mbox{regular terms},
\end{split}
\end{equation}
where the expression "regular terms", mean terms being regular at the 1-particle pole of the 2-point 
function.  

Using the continuum normalization of states, both $\Gamma_z$ and the product of matrix elements in (\ref{Tubb}) 
contain ill defined terms. Thus, regularization of the Dirac-delta terms is necessary. 
In this paper, we will use the regularization method of \cite{BBCL}. Namely, we regularize the 
Dirac-delta function by the formula:
\begin{equation} \label{dreg}
\begin{split}
\delta(x)\to \delta_\epsilon(x)=\frac{i}{2 \pi} \left(\frac{1}{x+i \,\epsilon} - \frac{1}{x-i\, \epsilon}\right),
\end{split}
\end{equation}
and we take the $\epsilon \to 0$ limit only at the end of the computations.
With this regularization, the Dirac-delta singularity becomes a pole singularity. 
Then these type of singularities can mix with the ones coming from the kinematical pole 
singularities of the 3-particle form-factors entering (\ref{m1}) and (\ref{m2}). 
 For any 3-particle form-factor $f$, it is useful to introduce the finite part, which is free from 
kinematical poles: 
\begin{equation} \label{f3c}
\begin{split}
f_{a b_1 b_2}^c(u,\beta_1,\beta_2)&=\!f_{a b_1 b_2}(u,\beta_1,\beta_2)-\frac{R_1[f]_{ab_1b_2}(\beta_1,\beta_2)}{u-\beta_1-i \,\pi}-
\frac{R_2[f]_{ab_1b_2}(\beta_1,\beta_2)}{u-\beta_2-i \, \pi},
\end{split}
\end{equation}
where the residues are of the form:
\begin{equation} \label{R1,2}
\begin{split}
R_1[f]_{a b_1 b_2}(\beta_1,\beta_2)&=\underset{\beta=\beta_1}{\mbox{Res}} \, f_{a b_1 b_2}(\beta\!+\!i \, \pi, \beta_1,\beta_2)\!=\!
\frac{i}{2 \pi}{\Big[} \delta_{\bar{a}b_1} \, f_{b_2}(\beta_2)\!-\!S_{b_1 b_2}^{\bar{a} x}(\beta_1\!-\!\beta_2) \,f_{x}(\beta_2)  {\Big]}, \\
R_2[f]_{a b_1 b_2}(\beta_1,\beta_2)&\!=\!\underset{\beta=\beta_2}{\mbox{Res}}\, f_{a b_1 b_2}(\beta\!+\!i \, \pi, \beta_1,\beta_2)\!=\!
\frac{(-1)^{F}}{2 \pi \, i}{\Big[} \delta_{\bar{a}b_2} \, f_{b_1}(\beta_1)\!-\!S_{b_1 b_2}^{x \bar{a}}(\beta_1\!-\!\beta_2) \,f_{x}(\beta_1)  {\Big]}.
\end{split}
\end{equation}
Furthermore, for later convenience it is also useful to introduce the form-factor, which is regular 
in the 1st pair of its arguments:
\begin{equation} \label{fhat}
\begin{split}
\hat{f}_{a b_1 b_2}(u,\beta_1,\beta_2)&=\!f_{a b_1 b_2}(u,\beta_1,\beta_2)-\frac{R_1[f]_{ab_1b_2}(\beta_1,\beta_2)}{u-\beta_1-i \pi}.
\end{split}
\end{equation}

Using these definitions, we introduce{\footnote{In analogy with the diagonally scattering case \cite{BBCL}.}}
 the regularized 3-particle form-factor by the definition as follows:
\begin{equation} \label{freg}
\begin{split}
f^{reg}_{a b_1 b_2}(u,\beta_1,\beta_2)&=\!\hat{f}_{a b_1 b_2}(u,\beta_1,\beta_2)+
\frac{i}{4 \pi} \, (S_{\bar{a} b_2}^{b_1 x})'(\beta_1-\beta_2)\, f_x(\beta_2),
\end{split}
\end{equation}
where $'$ denotes the derivative and for repeated indexes summation is meant.
At the end of this section, we will see, that this combination of form-factors 
and S-matrix elements will play a central 
role in our final formula for the L\"uscher-corrections of the 1-particle form-factors.

We just note, that definitions (\ref{f3c})-(\ref{freg}) are the natural generalizations 
of the analogous definitions of \cite{BBCL} in diagonally scattering integrable theories.

\subsection{Computing $J_{12}(u|\psi,q)[F,f]$}

The function $J_{12}(u|\psi,q)[F,f]$ is defined in (\ref{J12}). Using the method of \cite{BBCL}, we compute it 
in the vicinity of the 1-particle pole. As, a first step we rephrase it for real values of $\omega,$ 
or equivalently $\psi.$ Then we perform the analytical continuation from the real axis towards the 
interesting point $\omega \to i {\cal E}(q).$
 
We start with introducing the new integration variables:
 \begin{equation} \label{bw}
\begin{split}
\beta_1=b+w, \qquad \beta_2=b-w,
\end{split}
\end{equation}
Then the $b$-integration can be evaluated:
\begin{equation} \label{J12butan}
\begin{split}
J_{12}(u|\psi,q)[F,f]=\int\limits_{-\infty}^{\infty} \! dw \, 
\frac{T(u|b_0(w)+w,b_0(w)-w)}{C(b_0(w),w)\, (C(b_0(w),w)-\cosh u -i \, \hat{q})}, 
\end{split}
\end{equation}
where
\begin{equation} \label{Cb0}
\begin{split}
C(b,w)=2 \cosh b \, \cosh w, \qquad b_0(w)=\mbox{arcsinh}\left(\frac{\sinh u +\sinh \psi}{2 \cosh w}\right).
\end{split}
\end{equation}
If we restrict ourselves to the domain $\cosh u<2,$ which is enough for the $O(e^{-\ell})$ order L\"uscher-corrections, 
we can avoid the branch cuts coming from the $\mbox{arcsinh}$-function. 
Though, the $w$-integration is regular along the real axis, 
as a consequence of the $\epsilon$-regularization of the Dirac-delta function (\ref{dreg}), 
pole singularities are present from $\epsilon$ order distance from the real axis. 
On the upper half plane, the position of this type of pole is given by the formula:
\begin{equation} \label{wplus}
\begin{split}
w_+=w_0+\epsilon \, w_1+O(\epsilon^2), \qquad 
w_0=\frac{u-\psi}{2}, \qquad w_1=\frac{i}{2} \frac{\cosh u +\cosh \psi}{\cosh \psi}.
\end{split}
\end{equation}
To get rid of the cumbersome $\epsilon$-dependence of the integrand,
 we rephrase (\ref{J12butan}), by shifting the integration contour: $w 
\to w+i \, \gamma,$ with $\gamma$ being a small positive number, but large enough to run 
the integration contour above these infinitesimally close poles. Due to the pole, this shift 
results residue terms from the pole $w_+$ of (\ref{wplus}). 
Then $J_{12}(u|\psi,q)[F,f]$ takes the new form:
\begin{equation} \label{J12resutan}
\begin{split}
J_{12}(u|\psi,q)[F,f]=J_{R12}(u|\psi,q)[F,f]+J_0(u|\psi,q)[F,f],
\end{split}
\end{equation}
where $J_{R12}(u|\psi,q)[F,f]$ denotes the residue terms, and $J_0(u|\psi,q)[F,f]$ stands for the 
$\gamma$-shifted integral{\footnote{It is important to keep in mind, 
that functions entering (\ref{J12resutan}) are functionals of the form-factors $F$ and $f,$ even though 
the explicit indication of this dependence is neglected in the subsequent formulas.}}. 
The latter is of the form: 
\begin{equation} \label{J0}
\begin{split}
J_0(u|\psi,q)=\!\!\int\limits_{"i \gamma"} \!\!\! dw \, \frac{1}{\nu(\nu-\cosh u -i \hat{q})} 
{\Big [} H_0(u|b_0+w,b_0-w)+\sum\limits_{\sigma=\pm} \frac{H_1^{(\sigma)}(b_0+w,b_0-w)}{u-b_0-w} +\\
+\sum\limits_{\sigma=\pm} \frac{H_2^{(\sigma)}(b_0+w,b_0-w)}{u-b_0+w} {\Big ]}_{a b_1 b_2}
{\Big [} G_0(u|b_0+w,b_0-w)+
\sum\limits_{\sigma=\pm} \frac{G_1^{(\sigma)}(b_0+w,b_0-w)}{u-b_0-w} +\\
+\sum\limits_{\sigma=\pm} \frac{G_2^{(\sigma)}(b_0+w,b_0-w)}{u-b_0+w} {\Big ]}_{b_2 b_1 a}, 
\end{split}
\end{equation}
if with the help of the definitions (\ref{m1}), (\ref{m2}), (\ref{dreg}) and (\ref{f3c}) 
one formally rewrites $\Pi$ and $\bar{\Pi}$ as follows: 
\begin{equation} \label{PIuj}
\begin{split}
\Pi_{a b_1 b_2}(u|\beta_1,\beta_2)[F]&=\Big[H_0(u|\beta_1,\beta_2)+
\sum\limits_{\sigma=\pm} \frac{H_1^{(\sigma)}(u|\beta_1,\beta_2)}{u-\beta_1+i \, \sigma \, \epsilon}
+\sum\limits_{\sigma=\pm} \frac{H_2^{(\sigma)}(u|\beta_1,\beta_2)}{u-\beta_2+i \, \sigma \, \epsilon} \Big]_{a b_1 b_2}, \\
\bar{\Pi}_{a b_1 b_2}(u|\beta_1,\beta_2)[f]&=\Big[G_0(u|\beta_1,\beta_2)+
\sum\limits_{\sigma=\pm} \frac{G_1^{(\sigma)}(u|\beta_1,\beta_2)}{u-\beta_1+i \, \sigma \, \epsilon}
+\sum\limits_{\sigma=\pm} \frac{G_2^{(\sigma)}(u|\beta_1,\beta_2)}{u-\beta_2+i \, \sigma \, \epsilon} \Big]_{b_2 b_1 a}.
\end{split}
\end{equation}
In (\ref{J0}), $b_0\equiv b_0(w)$ given in (\ref{Cb0}), $\nu$ is defined in (\ref{omeganu}) 
and the integration contour runs along a straight line with 
distance $i \gamma$ above the real axis.
The $H[F]$ and $G[f]$-functions can be found in appendix \ref{appA}. 
After a lengthy computation the residue terms take the form as follows:
\begin{equation} \label{RJ12}
\begin{split}
J_{R12}(u|\psi,q)[F,f]=J_{R1}(u|\psi,q)+J_{R2}(u|\psi,q)+J_{R3}(u|\psi,q)+J_{R4}(u|\psi,q),
\end{split}
\end{equation}
where
\begin{equation} \label{R12}
\begin{split}
J_{R1}(u|\psi,q)=\frac{1}{\cosh \psi (\cosh \psi-i \hat{q})}\bigg[ \frac{i \, s_f}{4 \pi}\frac{{\cal K}_4}{u-\psi}+
s_f\, \sum\limits_x f_x(\psi)\, F_{\bar{x}}(\psi)\, \Big( \frac{s_0  \, n_0}{2\pi\epsilon}+
\delta(u-\psi)\Big)\bigg],
\end{split}
\end{equation}
\begin{equation} \label{R34}
\begin{split}
J_{R2}(u|\psi,q)&=\frac{i \, s_0 \, s_f {\cal K}_1}{4 \pi \,\cosh^2 \psi (\cosh \psi-i \hat{q})}
\bigg[\sinh(u-\psi)\, \bigg( \frac{1}{\cosh \psi-i \, \hat{q}}+\frac{1}{\nu} \bigg)-\frac{\Omega}{\nu \cosh \psi} \bigg]+\\
&+\frac{i \, s_0 \, s_f {\cal K}_3 \, \nu}{4 \pi \,\cosh^2 \psi (\cosh \psi-i \hat{q})},
\end{split}
\end{equation}
\begin{equation} \label{Rj2}
\begin{split}
J_{R3}(u|\psi,q)&=\frac{\sum\limits_{a,x} 
\Big[s_f \, f_{\bar{a} x a}^c(u+i \pi,\psi,u)+s_0 \, s_f f_{a \bar{a} x}(u+i \pi,u, \psi) \Big] \, F_{\bar{x}}(\psi)}
{2 \cosh \psi \, (\cosh \psi -i \, \hat{q})},
\end{split}
\end{equation}
\begin{equation} \label{Rj3}
\begin{split}
J_{R4}(u|\psi,q)&=\frac{\sum\limits_{a,x} 
\Big[s_f \, F_{\bar{a} x a}^c(u+i \pi,\psi,u)+s_0 \, s_f f_{\bar{a} a x}(u+i \pi,u, \psi) \Big] \, f_{\bar{x}}(\psi)}
{2 \cosh \psi \, (\cosh \psi -i \, \hat{q})},
\end{split}
\end{equation}
where
\begin{equation} \label{cK1}
\begin{split}
{\cal K}_1=\sum\limits_{a,x,y} F_{x}(\psi) 
\big[ S_{xa}^{ya}(\psi-u)-S_{ax}^{ay}(u-\psi)\big] f_{\bar{y}}(\psi),
\end{split}
\end{equation}
\begin{equation} \label{cK3}
\begin{split}
{\cal K}_3&=\sum\limits_{a,y,x} {\bigg\{} F_x(\psi) \big[(S_{xa}^{ya})'(\psi-u)+(S_{ax}^{ay})'(u-\psi) \big] 
f_{\bar{y}}(\psi)+\\+\frac{\cosh u}{\nu}
\! & \bigg(\! 
F_x'(\psi) \!\big[S_{xa}^{ya}(\psi\!-\!u)\!-\!S_{ax}^{ay}(u\!-\!\psi) \big] f_{\bar{y}}(\psi)\!+\!
F_x(\psi)\! \big[S_{xa}^{ya}(\psi\!-\!u)\!-\!S_{ax}^{ay}(u\!-\!\psi) \big] \!f_{\bar{y}}'(\psi)
\bigg)\!{\bigg\}},
\end{split}
\end{equation}
\begin{equation} \label{cK4}
\begin{split}
{\cal K}_4=\sum\limits_{a,x,y} {\bigg\{} F_{x}(u)\big[S_{xa}^{ay}(u\!-\!\psi)\!-\!S_{ax}^{ya}(\psi\!-\!u) \big] f_{\bar{y}}(\psi)+ 
F_{x}(\psi)\big[S_{ax}^{ya}(u\!-\!\psi)\!-\!S_{xa}^{ay}(\psi\!-\!u) \big]\! f_{\bar{y}}(u){\bigg\}},
\end{split}
\end{equation}
while
\begin{equation} \label{omeganu}
\begin{split}
\Omega=(\sinh u+\sinh \psi)\, (1+\sinh u \, \sinh \psi), \qquad 
\nu=\cosh u+\cosh \psi.
\end{split}
\end{equation}
Each of these formulas should be considered as functionals of the form-factors $f$ and $F,$ but to save space 
we do not indicated this dependence. As (\ref{GM12}) shows, to compute the 2-point function, the contribution 
of the simultaneously $f \to F^P, \quad F \to f^P$ and $q \to -q$ transformed $J_{12}$ will enter the 
calculations, as well.

So far, the building blocks of $\Gamma(\omega,q)$ have been computed for real values of $\omega$ 
or equivalently, for real $\psi.$ Now, we are in the position to discuss some trivial 
cancellations in our formulas. 
One can see, that there are terms being proportional to $\tfrac{1}{\epsilon}$ and $\delta(u-\psi)$ 
in (\ref{R12}) of (\ref{RJ12}). If according to (\ref{GMlus}), one adds all contributions to $\Gamma(\omega,q),$ 
then both the $\sim \delta(u-\psi)$  and $\sim \delta(0)$ terms cancel, if one makes the 
identification:
\begin{equation} \label{delta0}
\begin{split}
\delta(0)\to \frac{1}{2 \pi \epsilon}.
\end{split}
\end{equation}
We have to note, that literally applying the definition (\ref{dreg}), $\delta(0)=\tfrac{1}{\pi \epsilon}$ 
should be taken. Nevertheless, it is obvious, that the final result cannot be divergent, as such 
the $\sim \delta(0)$ terms should cancel, which can be reached by the identification (\ref{delta0}) for 
the ambiguous $\delta(0)$ value. Though, this prescription seems to be ad hoc, we use it, since it proved to 
be correct for diagonally scattering integrable theories \cite{BBCL}. There, it has been shown, that this
regularization of the $\delta$-function and $\delta(0)$ is equivalent to using finite volume 
regularization along the continuous direction. 

Now, one can conclude, that in the L\"uscher-correction for $\Gamma(\omega,q),$ the terms $-\Gamma_z \, \Gamma_{BY},$ 
and $\Gamma_{10}$ are present only to cancel the Dirac-delta terms arising in $\Gamma_{12}.$   

Thus, to get the L\"uscher-corrections close to the 1-particle pole, only $\Gamma_{12}(\omega,q)$ should be 
investigated, such that the $\tfrac{1}{\epsilon}$ and $\delta(u-\psi)$ terms are dropped from them. 

As a next step, we perform the analytical continuation in $\omega$ from the real axis to the value 
$i {\cal E}(q).$ 

\subsection{Analytical continuation}

As a first step, we introduce a new notation. Denote:
\begin{equation} \label{tJ12}
\begin{split}
\tilde{J}_{12}(u|\psi,q)[F,f]="{J}_{12}(u|\psi,q)[F,f] \, \mbox{without the } \tfrac{1}{\epsilon} \, 
\mbox{and} \, \delta(u-\psi) \, \mbox{terms.}"
\end{split}
\end{equation}
In this subsection, starting from the real axis, we perform the analytical continuation $\omega \to i {\cal E}(q).$ 
The residue terms listed in (\ref{RJ12})-(\ref{omeganu}), are analytic expressions in $\omega,$ thus 
their continuation is straightforward. On the other hand the shifted integral (\ref{J0}) requires more care.  

As one moves with $\omega$ from the real axis towards $i {\cal E}(q),$ according to (\ref{param}) 
the variable $\psi$ will evolve negative imaginary part, such that at $\omega=i {\cal E}(q):$
\begin{equation} \label{psiE}
\begin{split}
\psi \to \pm \theta-i \frac{\pi}{2}, \qquad \mbox{with} \quad q=m\sinh \theta.
\end{split}
\end{equation}
As a consequence, the pole of the integrand at 
$w_0=\tfrac{u-\psi}{2}$ tends to cross the integration contour. This crossing is taken into account by 
pushing the integration contour back to the real axis, such that the pole gives an additional residue term. 
After a tedious and lengthy computation, one ends up with the following result:
\begin{equation} \label{contres}
\begin{split}
\tilde{J}_{12}(u|\psi,q)[F,f]=\tilde{J}_{0}(u|\psi,q)[F,f]+\tilde{J}_{R}(u|\psi,q)[F,f],
\end{split}
\end{equation}
where  $\tilde{J}_{0}(u|\psi,q)[F,f]$ is the same as (\ref{J0}), but with integration along the real axis, 
and $\tilde{J}_{R}(u|\psi,q)[F,f]$ is the residue contribution. It can be easily shown, that $\tilde{J}_0$ 
doesnot have any poles at $\omega \sim i {\cal E}(q),$ thus this term doesnot contribute to the L\"uscher-correction 
of the 1-particle energies and form-factors. Only the residue term will give contributions to these quantities. 
They take the form as follows: 
\begin{equation} \label{RT}
\begin{split}
\tilde{J}_{R}(u|\psi,q)[F,f]=\frac{1}{\cosh \psi (\cosh \psi-i \hat{q})} \bigg( 
s_0 \, s_f \, \hat{W}_0+\frac{i \, s_f \, \hat{Y}_0}{2 \pi (u-\psi)}+
\frac{i \, s_0 \, s_f\, \nu \, \hat{\cal K}_3}{2 \pi \cosh \psi} \bigg)+\\
-\frac{i \, s_0 \, s_f\, \hat{\cal K}_1}{2 \pi \cosh^2 \psi (\cosh \psi-i \hat{q})}
\bigg( \sinh(u-\psi) \bigg(\frac{1}{\cosh \psi-i \hat{q}}+\frac{1}{\nu} \bigg)-
\frac{\Omega}{\nu \, \cosh \psi} \bigg),
\end{split}
\end{equation} 
where
\begin{equation} \label{hW0}
\begin{split}
\hat{W}_0(u,\psi,q)[F,f]=\sum\limits_{a,x} \big[
F^c_{\bar{a} a x}(u+i \pi,u,\psi) \, f_{\bar{x}}(\psi)+f^c_{a\bar{a} \bar{x}}(u+i\, \pi,u,\psi)\,
F_x(\psi) \big],
\end{split}
\end{equation}
\begin{equation} \label{hY0}
\begin{split}
\hat{Y}_0(u,\psi,q)[F,f]=\sum\limits_{x,y,a} 
\Big\{ F_x(u)  \big[ S_{xa}^{ay}(u\!-\!\psi)\!-\!\delta_{xy} \big] f_{\bar{y}}(\psi)\!+\!
 F_x(\psi)  \big[ S_{ax}^{ya}(u\!-\!\psi)\!-\!\delta_{xy} \big] f_{\bar{y}}(u) \Big\},
\end{split}
\end{equation}
\begin{equation} \label{cK1}
\begin{split}
\hat{\cal K}_1(u,\psi,q)[F,f]=\sum\limits_{x,y}  
F_x(\psi)  \sum\limits_a \big[ S_{ax}^{ay}(u\!-\!\psi)\!-\!\delta_{xy} \big] f_{\bar{y}}(\psi),
\end{split}
\end{equation}
\begin{equation} \label{cK3}
\begin{split}
\hat{\cal K}_3(u,\psi,q)[F,f]\!=\!\sum\limits_{a,x,y}\!  \bigg\{
F_x(\psi) (S_{ax}^{ay})'(u\!-\!\psi) f_{\bar{y}}(\psi)\!&-\!\tfrac{\cosh u}{\nu} 
F'_x(\psi)  \big[ S_{ax}^{ay}(u\!-\!\psi)\!-\!\delta_{xy} \big] f_{\bar{y}}(\psi)\\
&-\!\tfrac{\cosh u}{\nu} 
F_x(\psi)  \big[ S_{ax}^{ay}(u\!-\!\psi)\!-\!\delta_{xy} \big] f'_{\bar{y}}(\psi) \bigg\}.
\end{split}
\end{equation}
With help of (\ref{GMlus}), (\ref{GM12}), (\ref{S12}), (\ref{tJ12}) and (\ref{contres}) 
the L\"uscher-correction to the 2-point function at the 1-particle poles can be written as:
\begin{equation} \label{GLres}
\begin{split}
\Gamma_L(\omega,q)=\!\!\frac{2 \pi}{m^2 L} \int\limits_{} du \, e^{-\ell \cosh u} 
\big( \tilde{J}_{R}(u|\psi,q)[f\!,\!F]\!+\!s_0 \tilde{J}_{R}(u|\psi,-q)[F^P\!,\!f^P]\big)\!+\!\mbox{regular terms.}
\end{split}
\end{equation}
According to (\ref{Gkifejt1}) and (\ref{QR}), one needs to compute the coefficient of the single and double pole terms of (\ref{GLres})  at $\omega=i {\cal E}(q),$ to obtain the L\"uscher-correction of the form-factors and the 
mass gap. 

This is the next step in our calculations. Partly, one can work in the language of the variable $\psi.$ 
The location of the 1-particle pole singularity is at 
\begin{equation} \label{psi0pm}
\begin{split}
\psi \to \psi_0^{\pm}=\pm \theta-i \frac{\pi}{2}, \quad \mbox{with} \quad q=m \sinh \theta, 
\quad {\cal E}(q)=m \cosh \theta. 
\end{split}
\end{equation}
In the actual computations, we exploit, that a function being analytic around $\psi_0^{\pm},$ can be expanded as:
\begin{equation} \label{}
\begin{split}
{\cal G}(\psi(\omega))={\cal G}(\psi_0^{\pm})-\frac{1}{m} \frac{{\cal G}'(\psi_0^{\pm})}{\cosh \psi_0^{\pm}}\,
\big(\omega-i \, {\cal E}(q) \big)+O\big((\omega-i \, {\cal E}(q) )^2\big),
\end{split}
\end{equation}
and we make the substitution: 
\begin{equation} \label{ompsi}
\begin{split}
\frac{1}{\cosh \psi \pm i \, \hat{q}} \to \frac{m^2 (\cosh \psi\mp i \hat{q})}{\omega^2+q^2+m^2}=
\frac{m^2 (\cosh \psi\mp i \hat{q})}{(\omega-i \,{\cal E}(q))(\omega+i \,{\cal E}(q))}.
\end{split}
\end{equation}
Furthermore, and more importantly, we exploit the earlier assumed simple Kronecker-delta structure 
in the flavor space for the 1-particle form factors:
\begin{equation} \label{fFdelta}
\begin{split}
F_x(\theta)=\delta_{x \bar{x}_0} F_{\bar{x}_0}(\theta), \qquad f_x(\theta)=\delta_{x x_0} f_{x_0}(\theta).
\end{split}
\end{equation}
which allows one to get rid of plenty of summations, and obtain simpler formulas. 
Here, we emphasize again that the flavor space assumption (\ref{fFdelta}) 
is not only a theoretical assumption to get nicer results, but it is actually the case for 
 most of the important models, for example: for the fermion fields in the Massive Thirring model,
 or for the $O(N)$ vector fields in the $O(N)$ non-linear sigma models. 
For future convenience, it is worth introducing the function:
\begin{equation} \label{Ufv}
\begin{split}
U(\theta)=\sum\limits_a \big[ S_{a x_0}^{a x_0}(\theta+i \tfrac{\pi}{2})-1\big],
\end{split}
\end{equation}
where the sum runs over flavors of the lowest lying particle multiplet. As a consequence of the properties 
(\ref{Bose})-(\ref{Real}) 
of the $S$-matrix, it is independent of the choice of $x_0.$ 
Then, the residue of the double pole in (\ref{Gkifejt1}) takes the form:
\begin{equation} \label{cR}
\begin{split}
{\cal R}(\theta)=-\frac{m^2}{{\cal E}^2(q) \, L} \, s_{F^P}\, f^P_{x_0}(\psi_0^{+}) \, F^P_{\bar{x_0}}(\psi_0^+) \,
\int\limits du \, e^{-\ell \cosh u} \, \cosh(u-\theta) \, U(u-\theta),
\end{split}
\end{equation}
with $\psi_0^+$ given in (\ref{psi0pm}).
Comparing (\ref{cR}) to (\ref{QR}), allows one to extract the L\"uscher-correction for the 1-particle energy gap. 
The only ingredient one needs, is the identity:
\begin{equation} \label{BYfF}
\begin{split}
s_{F^P} \, f^P_{x_0}(\psi_0^+) \, F^P_{\bar{x}_0}(\psi_0^+)\big|_{BY}=
(-1)^F\, \frac{{\cal E}(q)\, L}{2 \pi} \,f^\psi_{x_0}(\theta) \, \bar{f}^{\bar{\psi}}_{\bar{x}_0}(\theta),
\end{split} 
\end{equation}
which can be derived by comparing the Bethe-Yang limit of (\ref{1partGMv1}) to the residue of  
 (\ref{GBY}) at the Bethe-Yang  limit of the 1-particle pole: $\omega=i {\cal E}(q)$ i.e. $\psi \to \psi_0^+,$ 
and exploiting the relation (\ref{BYinf}) 
between the Bethe-Yang and infinite volume limits of the 1-particle form factors.
Then, the L\"uscher-correction for the 1-particle energy gap can be written as:
\begin{equation} \label{dE}
\begin{split}
\delta {\cal E}(q)=-\frac{m}{\cosh \theta} \, \int\limits \frac{du}{2 \pi} \,
e^{-\ell \cosh u} \, \cosh(u-\theta) \, U(u-\theta), 
\end{split}
\end{equation}
which agrees with the result expected from the literature \cite{Luscher1,Luscher2,KlMe,JaLu,BaJa}.

The next step is to compute the corrections to the form-factors. As it was explained 
in section \ref{sect2}, the mirror framework is used to compute the L\"uscher-correction to 
the 1-particle form-factors. As a result, all formulas of the calculations contain the 
form-factors of the mirror transformed fields in the mirror theory. Certainly, at the end 
one would like to have formulas containing the form-factors of the original fields{\footnote{Here, the 
expression "original fields" mean, the fields, the finite volume form-factors of which we are interested in.}} 
in the original theory. Thus, to eliminate the mirror form-factors of the mirror-transformed 
fields, it is worth discussing the relation between the form-factors of the original theory
 and those of the mirror one.

As we will see in section \ref{sect5} in the example of the massive Thirring model, 
as a consequence of the relativistic invariance, in a relativistic theory the mirror theory 
can be made identical to the original one, and only the transformation of the fields 
under the mirror transformation should be taken into account. This is why, to eliminate 
the mirror form-factors from the final results for the L\"uscher-correction to the 
1-particle form-factors, only  
the knowledge of the relation between the form-factors of the original and 
mirror-transformed fields in the original theory becomes necessary. 

 In this respect, we make the following statement for the fundamental particle creating or annihilating 
boson or fermion fields of the original theory.  
Let $\psi$ such an operator with definite Lorentz-spin $s\in\{0,\pm 1/2\}$, and denote $\phi$ its image in 
the mirror theory (cf. (\ref{mirtf})). Then, their multiparticle form-factors are related by the formula:
\begin{equation} \label{Fconn}
\begin{split}
F^\psi_{a_1,...,a_n}(\theta_1,...,\theta_n)=F^{\phi^{\cal P}}_{a_1,...,a_n}(\theta_1-i \frac{\pi}{2},...,\theta_n-i\frac{\pi}{2}),
\end{split}
\end{equation}
which implies at the level of operators the following relation:
\begin{equation} \label{oprel}
\begin{split}
\psi(0)=e^{-i\tfrac{\pi}{2} s_{\phi^{\cal P}}} \, \phi^{\cal P}(0), \quad \mbox{with} \quad
\phi^{\cal P}(0)={\cal P}\phi(0){\cal P}^{-1},
\end{split}
\end{equation}
where ${\cal P}$ is the parity operator in the mirror theory, $s_{\phi^{\cal P}}$ is the spin of the operator 
$\phi^{\cal P},$ and in (\ref{Fconn}) in the left and right hand sides, the form-factors are meant in the 
original theory. 

The validity of formula (\ref{Fconn}) directly follows from (\ref{ax1}) for a spinless fundamental boson 
field in a parity invariant theory. Moreover, it will be clear from the subsequent sections, through the 
example of the massive Thirring model, that formulas (\ref{omg}), (\ref{mirmez0}) and 
(\ref{parity}) tell us, that (\ref{Fconn}) holds also for the fundamental fermion fields in a relativistically 
invariant fermion theory with quartic interactions{\footnote{We note, that based on some heuristic train of thoughts, we think that the form-factor connection 
formula (\ref{Fconn}) holds for arbitrary local operator with definite Lorentz-spin and not only for the fundamental 
fermion and boson fields.}}.

We note, that the identity (\ref{BYfF}), could also have been derived with the help of (\ref{BYinf}) and 
the form-factor relating formula (\ref{Fconn}).

Now, we are in the position to formulate the L\"uscher-corrections for the 1-particle form-factors. 
During the calculations, always the form-factors of the mirror image operators ($\phi,\bar{\phi}$)
 and of their parity transforms arise. Now, at the final stage, with the help of (\ref{Fconn}) 
the form-factor corrections can be expressed in terms of the form-factors of the original operators. 
To save space, we provide the final results only in this way. The L\"uscher-corrections of 
the 1-particle form-factors take the form as follows:
\begin{equation} \label{dfapsi}
\begin{split}
\delta f^\psi_x(\theta)=\frac{1}{\sqrt{\rho_1(\theta)}} {\bigg\{} 
\!-\!\frac{F^\psi_x(\theta)}{4 \pi} \bigg[ \frac{\sinh \theta}{\cosh^2 \! \theta} \Omega^{(0)}(\theta)+
\frac{\Omega^{(1)}(\theta)}{\cosh \theta}\bigg]+
\frac{{F^{\psi\prime}_x}(\theta)}{2 \pi} \frac{\Omega^{(0)}(\theta)}{\cosh \theta}+
\Omega^{\psi}_x(\theta){\bigg\}},
\end{split}
\end{equation}
\begin{equation} \label{dfbarapsi}
\begin{split}
\delta \bar{f}^{\bar{\psi}}_x(\theta)=\frac{1}{\sqrt{\rho_1(\theta)}} {\bigg\{} 
\!-\!\frac{\bar{F}^{\bar{\psi}}_x(\theta)}{4 \pi} \bigg[ \frac{\sinh \theta}{\cosh^2 \! \theta} \Omega^{(0)}(\theta)+
\frac{\Omega^{(1)}(\theta)}{\cosh \theta}\bigg]+
\frac{{\bar{F}^{\bar{\psi}\prime}_x}(\theta)}{2 \pi} 
\frac{\Omega^{(0)}(\theta)}{\cosh \theta}+
\Omega^{\bar{\psi}}_x(\theta){\bigg\}},
\end{split}
\end{equation}
where the notations for the form-factors (\ref{fBYjel}) and (\ref{fvegtelen}) are used 
and we introduced the short notations
\begin{equation} \label{OM0}
\begin{split}
\Omega^{(0)}(\theta)=\int\limits_{} \! du \, e^{-\ell \cosh u} \,\sinh u \, U(u-\theta),
\end{split}
\end{equation}
\begin{equation} \label{OM1}
\begin{split}
\Omega^{(1)}(\theta)=\int\limits_{} \! du \, e^{-\ell \cosh u} \, \sinh u \, U'(u-\theta),
\end{split}
\end{equation}
\begin{equation} \label{OMpsi}
\begin{split}
\Omega^{\psi}_x(\theta)=e^{i \tfrac{\pi}{2} s_\psi}\,\int\limits_{} \! du \, e^{-\ell \cosh u} \, \sum\limits_{a} 
F^{\psi,reg}_{a \bar{a} x}(u+i \pi,u,\theta-i \tfrac{\pi}{2}),
\end{split}
\end{equation}
\begin{equation} \label{OMbarpsi}
\begin{split}
\Omega^{\bar{\psi}}_x(\theta)=(-1)^F e^{i \tfrac{3 \pi}{2} s_{\bar{\psi}}}\,\int\limits_{} \! du \, e^{-\ell \cosh u} \, \sum\limits_{a} 
F^{\bar{\psi},reg}_{a \bar{a} \bar{x}}(u+i \pi,u,\theta-i \tfrac{\pi}{2}),
\end{split}
\end{equation}
where $s_\psi$ and $s_{\bar{\psi}}$ are the spins of the operators $\psi$ and $\bar{\psi},$ respectively,  
and the definition of the regularized form-factors entering (\ref{OMpsi}) and (\ref{OMbarpsi}) 
is given in (\ref{freg}), and $U(\theta)$ is defined in (\ref{Ufv}).

As it has been done in the diagonally scattering theories \cite{BBCL}, these results can be written in a more 
compact form. One can recognize, that the terms proportional to $\Omega^{(0)}$ and $\Omega^{(1)}$ 
in (\ref{dfapsi}) and (\ref{dfbarapsi}), can be reinterpreted as the L\"uscher-corrections of the 
1-particle density $\rho_1$ and of the rapidity. 
The L\"uscher-correction for 1-particle rapidity is given by the Bajnok-Janik formula \cite{BaJa}:
\begin{equation} \label{dtheta}
\begin{split}
\delta \theta=\frac{\Omega^{(0)}(\theta)}{2 \pi \cosh \theta}+O(e^{-2 \ell}).
\end{split}
\end{equation}
The density of states also gets exponentially small corrections of the form:
\begin{equation} \label{drho1}
\begin{split}
\delta\!\rho_1(\theta)=\rho_1(\theta) \, \frac{\Omega^{(1)}(\theta)}{2 \pi \cosh \theta}+O(e^{-2 \ell}).
\end{split}
\end{equation}
Denote $\rho(\theta)$ the exact 1-particle density, including all finite volume 
corrections{\footnote{ 
In the exact integrable description of the sine-Gordon/massive Thirring model $\rho(\theta)=\tfrac{Z'(\theta)}{2\pi}$ 
with $Z(\theta)$ being the counting-function in the NLIE description. For more details see: \cite{FevPhd} and references therein.}}:
\begin{equation} \label{rho}
\begin{split}
\rho(\theta)=\rho_1(\theta)+\delta\rho_1(\theta).
\end{split}
\end{equation}  
Then, rephrasing (\ref{dfapsi}) and (\ref{dfbarapsi}), the 1-particle form-factors 
upto L\"uscher-order can be written as follows:   
\begin{equation} \label{fapsi}
\begin{split}
f^{\psi}_a(\theta)+\delta f^{\psi}_a(\theta)=
\frac{F^\psi_x(\theta+\delta\theta)}{\sqrt{\rho(\theta+\delta\theta)}}
+\Omega^{\psi}_x(\theta)+O(e^{-2\ell}),
\end{split}
\end{equation}
\begin{equation} \label{fbarapsi}
\begin{split}
\bar{f}^{\bar{\psi}}_a(\theta)+\delta \bar{f}^{\bar{\psi}}_a(\theta)=
\frac{\bar{F}^{\bar{\psi}}_x(\theta+\delta\theta)}{\sqrt{\rho(\theta+\delta\theta)}}+
\Omega^{\bar{\psi}}_x(\theta)+O(e^{-2\ell}),
\end{split}
\end{equation}
Thus, the form-factor correction is composed of two type of terms. The 1st one is coming from 
taking the leading expression at the exact Bethe rapidity and with the exact density of states. 
The second term is a new type of term. It contains the higher multiparticle form-factors, corresponding 
to the contribution of a virtual particle traveling around the word.

\section{The Massive Thirring-model} \label{sect5}

In this paper, we will check our final results (\ref{fapsi}) and (\ref{fbarapsi}) for the 
1-particle form-factors through a perturbative calculation in the massive Thirring model.
In this section, we introduce the model and recall its most important properties. Moreover, we 
demonstrate, how the mirror transformation acts on the fields (c.f (\ref{mirtf})).
 
We start with the action of the Minkowski theory. Since at a later stage, we need to determine 
 the low order perturbative part of the 1- and 3-particle form-factors, 
it is worth using the conventions{\footnote{Our Lagrangian slightly differs from that of ref. \cite{BABK1}, 
as a consequence of a $\bar{\psi}\to -\bar{\psi}$ transformation.}} 
of ref. \cite{BABK1}:  
\begin{equation} \label{Smink}
\begin{split}
S[\bar{\psi},\psi]=\int\limits \! dt dx \, {\big\{} 
-\bar{\psi}\big(i \, \Gamma^0 \partial_t+i \Gamma^1 \partial_x-m \big)\psi
-\tfrac{g}{2} \, \bar{\psi}\Gamma^\mu \psi \, \bar{\psi}\Gamma_\mu \psi
{\big\}},
\end{split}
\end{equation}
where the $\Gamma$-matrices satisfy the algebra:
\begin{equation} \label{Galg}
\begin{split}
{\big{\{}} \Gamma^\mu,\Gamma^\nu {\big{\}}}=2 \, \eta^{\mu \nu},
\qquad {\big{\{}} \Gamma_\mu,\Gamma_\nu {\big{\}}}=2 \, \eta_{\mu \nu},
\qquad \mu,\nu=0,1,
\end{split}
\end{equation}
with $\eta$ being the metric tensor with the components:
\begin{equation} \label{eta}
\begin{split}
\eta^{\mu \nu}=\eta_{\mu \nu}=\mbox{diag}(1,-1).
\end{split}
\end{equation}
For the form-factor computations, we will use the chiral representation as follows:
\begin{equation} \label{Galak}
\begin{split}
\Gamma^0=\begin{pmatrix} 0 & 1 \\ 1 & 0
\end{pmatrix}, \qquad 
\Gamma^1=\begin{pmatrix} 0 & 1 \\ -1 & 0
\end{pmatrix}, \qquad 
\Gamma^5=\Gamma^0\Gamma^1\begin{pmatrix} -1 & 0 \\ 0 & 1
\end{pmatrix}.
\end{split}
\end{equation}
As for the fermion fields, they are two component vectors in the spinor space. The upper 
component will be denoted by $+$ and the lower one by $-$ subscripts.
\begin{equation} \label{psi}
\begin{split}
\psi=
\begin{pmatrix} \psi_+ \\ \psi_-
\end{pmatrix}, \qquad 
\bar{\psi}=
\begin{pmatrix} \bar{\psi}_+ \\ \bar{\psi}_-
\end{pmatrix}.
\end{split}
\end{equation}
In our notations $\bar{\psi}$ is related to $\psi$ by the formula: $\bar{\psi}=-\psi^+ \, \Gamma^0.$
The Euclidean action can be obtained from the Minkowskian one by the usual Wick-rotation: $t \to -i \, \tau:$
\begin{equation} \label{Seucl}
\begin{split}
S_E[\bar{\psi},\psi]=\int\limits \! d\tau dx \, {\big\{} 
-\bar{\psi}\big( \gamma_0 \partial_\tau+ \gamma_1 \partial_x+m \big)\psi
+\tfrac{g}{2} \, \bar{\psi}\gamma_\mu \psi \, \bar{\psi}\gamma_\mu \psi
{\big\}},
\end{split}
\end{equation}
where now $\gamma_\mu$ denotes the Euclidean $\gamma$-matrices satisfying the algebra
\begin{equation} \label{galg}
\begin{split}
{\big{\{}} \gamma_\mu,\gamma_\nu {\big{\}}}=2 \, \delta_{\mu \nu}, \qquad \mu,\nu=0,1.
\end{split}
\end{equation}
They are simply related to their Minkowskian counterpart given in (\ref{Galak}):
\begin{equation} \label{gG}
\begin{split}
\gamma_0=\Gamma^0, \qquad \gamma_1=-i\, \Gamma^1.
\end{split}
\end{equation}
The mirror transformation exchanges the role of space and time. As a consequence  
the two $\gamma$-matrixes change place, resulting the mirror action:
\begin{equation} \label{SMeucl}
\begin{split}
S_M[\bar{\phi},\phi]=\int\limits \! d\tau dx \, {\big\{} 
-\bar{\phi}\big( \tilde{\gamma}_0 \partial_\tau+ \tilde{\gamma}_1 \partial_x+m \big)\phi
+\tfrac{g}{2} \, \bar{\phi}\tilde{\gamma}_\mu \phi \, \bar{\phi}\tilde{\gamma}_\mu \phi
{\big\}},
\end{split}
\end{equation}
where $\phi$ and $\bar{\phi}$ are the fundamental fermion fields and the $\tilde\gamma$-
matrices are the $\gamma$-matrices in the mirror theory. They also satisfy the usual algebra:
\begin{equation} \label{gtalg}
\begin{split}
{\big{\{}} \tilde\gamma_\mu,\tilde\gamma_\nu {\big{\}}}=2 \, \delta_{\mu \nu}, \qquad \mu,\nu=0,1,
\end{split}
\end{equation}
and they are related to the "original ones" by a simple exchange in their subscripts:
\begin{equation} \label{gk}
\begin{split}
\tilde\gamma_0=\gamma_1, \qquad \tilde\gamma_1=\gamma_0.
\end{split}
\end{equation}
This exchange can be rephrased more elegantly by a unitary transformation $\omega_\gamma$ satisfying:
\begin{equation} \label{omgrel}
\begin{split}
\omega_\gamma \, \gamma_\mu \omega_\gamma^+=\tilde\gamma_\mu, \qquad \mu=0,1.
\end{split}
\end{equation}
This fixes $\omega_\gamma$ upto a sign. The solution of (\ref{omgrel}), we will use in the sequel, is as follows: 
\begin{equation} \label{omg}
\begin{split}
\omega_\gamma=
\begin{pmatrix} 0 & e^{-\tfrac{i \, \pi}{4}} \\
e^{\tfrac{i \, \pi}{4}} & 0
\end{pmatrix}.
\end{split}
\end{equation}
It has the simple properties as follows:
\begin{equation} \label{omgprop}
\begin{split}
\omega_\gamma=\omega_\gamma^+=\omega_\gamma^{-1}, \qquad \omega_\gamma^*=\omega_\gamma^T,
\end{split}
\end{equation}
where $*$ and $T$ mean complex conjugation and transposition, respectively.
The matrix $\omega_\gamma$ allows one to determine, how the fields transform under the mirror 
transformation. One can show that the two Euclidean actions (\ref{Seucl}) and  (\ref{SMeucl}) 
are identical, if the following relations among the fields are assumed: 
\begin{equation} \label{mirmez}
\begin{split}
\phi(\tau,x)=\omega_\gamma\, \psi(x,\tau), \qquad 
\bar{\phi}(\tau,x)= \bar{\psi}(x,\tau) \, \omega_\gamma^+.
\end{split}
\end{equation}
One can recognize, that apart from the linear transformation mixing the components, 
the exchange of arguments is also important. Since the space-time dependence of the 
fields is simple and well known, from the point of view of our form-factor computations, 
this relation becomes more important at the origin, 
where it reduces to a simple linear transformation of the components: 
\begin{equation} \label{mirmez0}
\begin{split}
\phi(0)=\omega_\gamma\, \psi(0), \qquad 
\bar{\phi}(0)= \bar{\psi}(0) \, \omega_\gamma^+.
\end{split}
\end{equation}
For short, denote the exchange of arguments in $\psi$ and $\bar{\psi}$ by a prime. Namely,
 \begin{equation} \label{pprime}
\begin{split}
\psi(x,\tau)=\psi'(\tau,x), \qquad \bar{\psi}(x,\tau)=\bar{\psi}'(\tau,x).
\end{split}
\end{equation}
This simplified notation allows us to discuss the propagator at the level of path integral.
\begin{equation} \label{prop}
\begin{split}
\langle \bar{\psi}(x,\tau){\psi}(0) \rangle_L=\frac{1}{Z}\!\!\int\limits_{{}_{"x+L=x "}}\!\! 
{\cal D}\bar{\psi} {\cal D}{\psi} \,\,  \bar{\psi}(x,\tau){\psi}(0) \, e^{-S_E(\bar{\psi},\psi)}= 
\qquad \qquad \qquad \quad \\
=\frac{1}{Z}\!\!\int\limits_{{}_{"\tau_m+L=\tau_m "}}\!\! 
{\cal D}\bar{\psi}' {\cal D}{\psi'} \,\,  \bar{\psi'}(\tau_m,x_m){\psi'}(0) \, e^{-S_M(\bar{\psi}',\psi')}=
\qquad \qquad \qquad \\
=\frac{1}{Z}\!\!\int\limits_{{}_{"\tau_m+L=\tau_m "}} \!\! 
{\cal D}\bar{\psi}' {\cal D}{\psi'} \,\,  
(\bar{\psi'}\omega_\gamma^+)(\tau_m,x_m)(\omega_\gamma \psi')(0) \, e^{-S_E(\bar{\psi}',\psi')}=
\langle \bar{\phi}(x_m,\tau_m){\phi}(0) \rangle_{T=\tfrac{1}{L}},
\end{split}
\end{equation}
where $\tau_m=x$ and $x_m=\tau$ denote the mirror time- and space- coordinates. 
In the 1st step we reinterpreted time and space through the mirror model, and in the 2nd step we 
changed integration variables $\bar{\psi}' \to \bar{\psi}'\omega_\gamma^+$ and 
${\psi}' \to \omega_\gamma \, \psi',$ exploiting the relation 
$S_M(\bar{\psi}' \, \omega_\gamma,\omega_\gamma \, \psi')=S_E(\bar{\psi}',\psi').$
Formula (\ref{prop}) implies, that the original finite volume propagator, can be identically 
represented as a finite temperature propagator with temperature $T=\tfrac{1}{L},$ in the same 
theory, but considering it as the 2-point thermal correlator of the transformed fields (\ref{mirmez}). 

In the previous section, we computed the L\"uscher-corrections of the 1-particle form-factors, 
starting form the finite volume propagator. The L\"uscher-corrections could be computed from the 
mirror channel, where the fields constituting the propagator are mirror transformed counterparts
 of the original fields. 
Thus, during the computation always the form-factors of these transformed fields in the mirror 
theory appear. In general, one might think, that the form-factors in another model being different 
from the original one, should be known to get the form-factor corrections. 
Nevertheless, formula (\ref{prop}) shows the commonly
 known fact, that in a relativistically invariant theory, the mirror theory is equivalent to the  
original one. Namely, in our example the mirror problem can be rephrased, such that the thermal correlators,
 of the transformed fields have to be computed, but in exactly the same theory as the original was. 
Consequently, to compute the L\"uscher-corrections to the form-factors, only the form-factors of the 
original theory will arise. This explains the fact, why the form-factors of the same 
theory can be found on both sides of the form-factor connection formula (\ref{Fconn}). 

In the rest of the paper we will check our formulas (\ref{fapsi}) and (\ref{fbarapsi}) 
for the L\"uscher-correction of 1-particle form-factors at leading order in perturbation theory. 
This requires two type of computations. On the one hand, with the help of integrability, 
one can determine the exact S-matrix and form-factors of the model. This makes 
it possible to determine the perturbative series of the form-factor correction formulas
 (\ref{fapsi}) and (\ref{fbarapsi}).  On the other hand, the 2-point function at the 1-particle 
pole can be directly computed from Lagrangian perturbation theory. The comparison of the results 
from the two different methods, allows us the give a nontrivial check on our formulas.

Now, we start with the computation on the integrable side.

\section{Weak coupling series from integrability} \label{sect6}

In this section we perform the weak coupling expansion of the formulas (\ref{fapsi}) and (\ref{fbarapsi})  
for the L\"uscher-correction to the 1-particle form-factors. 
As a starting point it is worth investigating the free theory.

\subsection{The free fermion case}

The investigation of the free case is important, since when solving the axioms (\ref{ax1})-(\ref{ax4})  
for the infinite volume form-factors, 
one has a freedom in the normalization of 1-particle form-factors.  
Our choice is that we fix them to the values coming from the free fermion theory. 
In the free Minkowski theory, the fermion fields admit the Fourier-expansion:
\begin{equation} \label{ppbarfft}
\begin{split}
\psi(x,t)&=\int\limits\!\frac{dq}{(4 \pi)^{3/2} {\cal E}(q)} \,\bigg( 
{\bf b}(q)\, u(q) \, e^{-i {\cal E}(q) t+i \, q x}+ {\bf d}^+(q)\, v(q)\, e^{i {\cal E}(q) t-i q x}
\bigg), \\
\psi^+(x,t)&=\int\limits\!\frac{dq}{(4 \pi)^{3/2} {\cal E}(q)} \,\bigg( 
{\bf b}^+(q)\, u^*(q) \, e^{i {\cal E}(q) t-i \, q x}+ {\bf d}(q)\, v^*(q)\, e^{-i {\cal E}(q) t+i q x}
\bigg),
\end{split}
\end{equation}
where $*$ stands for complex conjugation, and the operators ${\bf b},{\bf b}^+,{\bf d},{\bf d}^+$ 
fermion and anti-fermion creation and annihilation operators with the anti-commutation relations:
\begin{equation} \label{bd}
\begin{split}
{\big\{}{\bf b}^+(q),{\bf b}(q') {\big\}}={\big\{}{\bf d}^+(q),{\bf d}(q') {\big\}}= {\cal E}(q) \,
\delta(q-q'), \\
{\big\{}{\bf b}(q),{\bf b}(q') {\big\}}={\big\{}{\bf b}^+(q),{\bf b}^+(q') {\big\}}=
{\big\{}{\bf d}(q),{\bf d}(q') {\big\}}={\big\{}{\bf d}^+(q),{\bf d}^+(q') {\big\}}=0.
\end{split}
\end{equation}
The $u$- and $v$-spinors in rapidity parameterization $q=m \sinh \theta$ take the forms \cite{BABK1}:
\begin{equation} \label{uv}
\begin{split}
u(\theta)=\sqrt{m}
\begin{pmatrix} e^{-{\theta}/{2}} \\ e^{{\theta}/{2}}
\end{pmatrix}, \qquad
v(\theta)= i\, \sqrt{m}
\begin{pmatrix} e^{-{\theta}/{2}} \\ -e^{{\theta}/{2}}
\end{pmatrix}.
\end{split}
\end{equation}
The operators, 
${\bf {b}}^+(q)$ and ${\bf {d}}^+(q)$  create the states with pure Dirac-delta normalization in the rapidity variable:  
\begin{equation} \label{btdt}
\begin{split}
|\theta\rangle_+={\bf {b}}^+(q)|0\rangle, \qquad 
|\theta\rangle_{-}={\bf {d}}^+(q)|0\rangle,
\end{split}
\end{equation}
where $+$ denotes the fermion and $-$ the anti-fermion in the flavor space. 

Thus, the infinite volume 1-particle form-factors in our normalization take the form:
\begin{equation} \label{1ffnorm}
\begin{split}
\langle 0| \psi(0)|\theta\rangle_a=F^\psi_a(\theta)=\delta_{a +} \frac{u(\theta)}{\sqrt{4 \pi}}, \\
\langle 0| \bar{\psi}(0)|\theta\rangle_a=F^{\bar\psi}_a(\theta)=\delta_{a -} \frac{\bar{v}(\theta)}{\sqrt{4 \pi}}.
\end{split}
\end{equation}
In the massive Thirring model, we normalize the infinite volume form-factors of elementary fermion 
 fields, such that the 1-particle form-factors are exactly given by their free fermion limits (\ref{1ffnorm}).
 
\subsection{Symmetries}

For the form-factor computations, it is worth summarizing, how the fields transform under 
parity and charge conjugation. A simple computation shows, that with the action (\ref{Smink}), the 
parity transformation of the fields takes the usual form:
\begin{equation} \label{parity}
\begin{split}
{\cal P} \, \psi(0) \, {\cal P}^{-1}=\Gamma^0\, \psi(0), \qquad 
{\cal P} \, \bar{\psi}(0) \, {\cal P}^{-1}=\bar{\psi}(0) \, \Gamma^0, 
\end{split}
\end{equation}
where ${\cal P}$ is the parity operator with the action on the 1-particle states as follows:
\begin{equation} \label{Pop}
{\cal P}|\theta\rangle_a=p_{a}|-\theta\rangle_a, \qquad \mbox{with} \quad 
p_a=\delta_{a +}-\delta_{a -}.
\end{equation}
Its action on a general multiparticle state can be given by:
\begin{equation} \label{Pmult}
\begin{split}
{\cal P}|\theta_1,...,\theta_n)_{a_1,... ,a_n}=(-1)^{n(n-1)/2} \,\prod\limits_{j=1}^n p_{a_j} |-\theta_n,..,-\theta_1 
\rangle_{a_n,..,a_1}.
\end{split}
\end{equation}

Another important transformation is the charge-conjugation symmetry, which transforms a particle 
to its anti-particle. Its action on 1-particle states is given in (\ref{Ckonj}), and it acts 
on a general multiparticle state as follows:
\begin{equation} \label{Cmult}
\begin{split}
C|\theta_1,...,\theta_n\rangle_{a_1,...,a_n}=|\theta_1,..., \theta_n \rangle_{\bar{a}_1,..,\bar{a}_n}.
\end{split}
\end{equation}
Its action on the fernion fields take the form:
\begin{equation} \label{Ctf}
\begin{split}
C \,\psi(0)\,C^{-1}=-i \, \Gamma^5 \, \psi^+(0)=-i \Gamma^1 \, \bar{\psi}(0), \qquad\quad 
C \,\psi^+(0)\,C^{-1}=i \, \Gamma^5 \, \psi(0), 
\end{split}
\end{equation}
with the $\bar{\psi}(0)=-\psi^+(0)\, \Gamma^0$ convention.
In the following subsections the transformation rules (\ref{Ctf}) will prove to be particularly useful, 
since they will allow to determine the matrix elements of the operators $\bar{\psi}$ and $\psi^+$ from  
those of  $\psi.$ 

\subsection{The S-matrix}

The knowledge of the two-particle S-matrix is essential to determine the form-factors 
of local operators through solving the axioms (\ref{ax1})-(\ref{ax4}). The S-matrix of the 
massive Thirring model has been known for a long time \cite{ZamiZami}. Denote $+$ the fermion and $-$ the anti-fermion 
state. Then, the non-zero S-matrix elements take the forms:
\begin{equation} \label{Sm}
\begin{split}
S_{++}^{++}(\theta)&=S_{--}^{--}(\theta)=a(\theta)=
\exp \int\limits_{0}^{\infty} \frac{dt}{t} 
\frac{\sinh \tfrac{(1-\nu)t}{2}}{\sinh \tfrac{\nu t}{2} \, \cosh \tfrac{t}{2}}
\, \sinh \tfrac{t\, \theta}{i \pi}, \\
S_{+-}^{+-}(\theta)&=S_{-+}^{-+}(\theta)=b_0(\theta) \, a(\theta), \quad \mbox{with} \quad
 b_0(\theta)=\frac{\sinh\tfrac{\theta}{\nu}}{\sinh\tfrac{i \, \pi -\theta}{\nu}}, \\
S_{+-}^{-+}(\theta)&=S_{-+}^{+-}(\theta)=c_0(\theta) \, a(\theta), \quad \mbox{with} \quad
 c_0(\theta)=\frac{\sinh\tfrac{i \, \pi}{\nu}}{\sinh\tfrac{i \, \pi -\theta}{\nu}},
\end{split}
\end{equation}
where the parameter $\nu$ carries the information on how the S-matrix depends on the
coupling constant of the Lagrangian (\ref{Smink}) of the model. 
This is given by the relation:
\begin{equation} \label{nug}
\begin{split}
\nu=\frac{\pi}{\pi+2 \, g}.
\end{split}
\end{equation}
In the sequel we focus on the regime $1<\nu,$ or equivalently $g<0,$  where 
there are no bound states, and only the fermion and the anti-fermion form
the particle spectrum of the model.

\subsection{The 3-particle form-factors}

To test our formulas (\ref{fapsi}) and (\ref{fbarapsi}) for the L\"uscher-correction of 1-particle form-factors, 
the 3-particle form-factors of the corresponding fields should be determined. We test our formula by the 
fermion propagator, thus we need to know the 3-particle form-factors of the fields $\psi$ and $\bar{\psi}.$
There are several ways in the literature to solve the form-factor axioms for our model. 
Beyond Smirnov's seminal work \cite{Smirnov}, the free-field representation \cite{LUK1,LUK2}, and the off-shell Bethe-Ansatz method \cite{BABK1,BABK2} proved to be very useful for the determination of form-factors in the sine-Gordon/massive 
Thirring model. In this paper we use the latter method for computing perturbatively the necessary 3-particle 
form-factors. To be pragmatic, we present only the most necessary formulas. For a detailed description,  
the reader is referred to the original paper \cite{BABK1}.  
What we compute from the off-shell Bethe-Ansatz method are the 3-particle form-factors of the operator $\psi.$ 
All the other necessary form-factors can be determined with the help of the charge conjugation transformations 
(\ref{Ctf}). 

\subsubsection{3-particle form-factors of $\psi_\pm(0)$}

In \cite{BABK1}, the 1-loop 3-particle form-factors of the operators $\psi_\pm(0)$ have been determined from 
the weak coupling expansion of the exact formula for the form-factors. Now, we shortly 
review the computation and present the 1-loop order result. 
 From \cite{BABK1}, the exact 3-particle form-factor admits the representation:
\begin{equation} \label{f3repr}
\begin{split}
f^{\psi_\pm}_{a_1 a_2 a_3}(\theta_1,\theta_2,\theta_3)=\hat{N}^{\psi_\pm}_3\!\!
\left[\!\!\prod\limits_{1<i<j\leq3} \!\!\!\!\!\!F(\theta_i\!-\!\theta_j) \right]\!
\int\limits_{C_{\underline{\theta}}} \! du \prod\limits_{i=1}^3 \phi(\theta_i\!-\!u)
 e^{\pm(u-\bar{\theta})}\, 
\Psi_{a_1 a_2 a_3}(u|\theta_1,\theta_2,\theta_3),
\end{split}
\end{equation}
where $\bar{\theta}=\frac{\theta_1+\theta_2+\theta_3}{2}$ and  
the Bethe-wave function $\Psi$ is defined through the matrix elements of a monodromy matrix:
\begin{equation} \label{Psi3}
\begin{split}
\Psi_{a_1 a_2 a_3}(u|\theta_1,\theta_2,\theta_3)&={\cal T}_{+,a_1 a_2 a_3}^{-,+++}(\theta_1,\theta_2,\theta_3|u), \\
{\cal T}_{a,a_1 a_2 a_3}^{b,b_1 b_2 b_3}(\theta_1,\theta_2,\theta_3|u)&=
\sum\limits_{x,y} \dot{S}_{a a_3}^{x b_3}(\theta_3-u) \, \dot{S}_{x a_2}^{y b_2}(\theta_2-u) \,
\dot{S}_{y a_1}^{b b_1}(\theta_1-u),
\end{split}
\end{equation}
where according to (\ref{Sdot}) $\dot{S}=-S$ for the massive Thirring model. The integration contour prescription $C_{\underline{\theta}}$ in (\ref{f3repr}) has the following action on an arbitrary function of $u:$
\begin{equation} \label{Ctheta}
\begin{split}
\int\limits_{C_{\underline{\theta}}} \! du \, f(u)&=2 \pi \, i \sum\limits_{j=1}^3 \!\!
\left( \underset{\theta_j-i \pi}{\mbox{Res}}\!-\!\frac{1}{2}\underset{\theta_j}{\mbox{Res}}
\!+\!\frac{1}{2}\underset{\theta_j+i \pi \, (\nu-1)}{\mbox{Res}}\right) f(u)\!+\!
\int\limits_{-\infty}^{\infty} \!\frac{du}{2} \bigg\{ f(u+i \tfrac{\pi}{2})+f(u-i \tfrac{\pi}{2})\bigg\}.
\end{split}
\end{equation}
In addition, $\hat{N}_3^{\psi_\pm}$ is a normalization factor with the definition,
\begin{equation} \label{N3}
\begin{split}
\hat{N}_3^{\psi_\pm}=\pm\,  \frac{i\, m^{1/2}}{(4 \pi)^{5/2}}\,f_{ss}^{min}(0)^2.
\end{split}
\end{equation}
The so far undefined functions in (\ref{f3repr}) are given by the integral representations as follows 
\cite{BABK1}:
\begin{equation} \label{fss}
\begin{split}
f_{ss}^{min}(\theta)&=\exp \int\limits_{0}^\infty \! \frac{dt}{t} 
\frac{\sinh \tfrac{(1-\nu)\,t}{2}}{\sinh\tfrac{\nu \,t}{2} \, \cosh \tfrac{t}{2}} \,
\frac{1-\cosh\left[ t \,(1-\theta/(i \, \pi))\right]}{2 \, \sinh t}, \\
F(\theta)&=-i\, \sinh \tfrac{\theta}{2} \, f_{ss}^{min}(\theta), \\
\phi(u)&=\frac{i}{F^2\left(\tfrac{i \pi}{2}\right)\, \sinh u}\, 
\exp \int\limits_{0}^\infty \! \frac{dt}{t} 
\frac{\sinh \tfrac{(1-\nu)\,t}{2}\, \left( \cosh \left[t \, (1/2-u/(i \pi))\right]-1\right)}{\sinh\tfrac{\nu \,t}{2} \, \sinh t}. 
\end{split}
\end{equation}
For further useful representations and recurrence relations see appendix C in \cite{BABK1}. 
The 1-loop evaluation of the 3-particle form-factor formula (\ref{f3repr}) can also be found in 
\cite{BABK1}.

To save space, we present only the 
$-++$ component, because the other nonzero elements can be determined from it simply by the application of 
the cyclic axiom (\ref{ax3}). 
\begin{equation} \label{1loopf3}
\begin{split}
f^{\psi_+}_{-++}(\theta_1,\theta_2,\theta_3)&=g\, \frac{i \sqrt{m} e^{\frac{1}{2} \left(\theta _1-\theta _2-\theta
   _3\right)} \left(e^{\theta _3}-e^{\theta _2}\right)
   \left(e^{\theta _3} \left(e^{\theta _1}+2 e^{\theta
   _2}\right)+e^{\theta _1+\theta _2}\right)}{4 \pi ^{3/2}
   \left(e^{\theta _1}+e^{\theta _2}\right) \left(e^{\theta
   _1}+e^{\theta _3}\right) \left(e^{\theta _2}+e^{\theta
   _3}\right)}+O(g^2),\\
f^{\psi_-}_{-++}(\theta_1,\theta_2,\theta_3)&=g\,\frac{i \sqrt{m} e^{\frac{1}{2} \left(\theta _1+\theta _2+\theta
   _3\right)} \left(e^{\theta _3}-e^{\theta _2}\right) \left(2
   e^{\theta _1}+e^{\theta _2}+e^{\theta _3}\right)}{4 \pi ^{3/2}
   \left(e^{\theta _1}+e^{\theta _2}\right) \left(e^{\theta
   _1}+e^{\theta _3}\right) \left(e^{\theta _2}+e^{\theta
   _3}\right)}+O(g^2).
\end{split}
\end{equation}
To compute the finite volume correction to the residue of the fermion propagator the form-factors of the 
fields $\bar{\psi}_\pm(0)$ are also required. They can be determined from those of $\psi_\pm(0),$ with the 
help of the charge conjugation symmetry (\ref{Ctf}), which implies:  
\begin{equation} \label{ppbarrel}
\begin{split}
{}_{b_1,..,b_m}\langle \theta_1',..,\theta_m'|\bar{\psi}_\pm(0)|\theta_1,..,\theta_n\rangle_{a_1,..,a_n}=
\mp i \,{}_{\bar{b}_1,..,\bar{b}_m}\langle \theta_1',..,\theta_m'|{\psi}_\pm(0)|\theta_1,..,\theta_n\rangle_{\bar{a}_1,..,\bar{a}_n} .
\end{split}
\end{equation}
Luckily, for the form-factor corrections one needs only special combinations of the 3-particle form factors 
(\ref{OMpsi}), (\ref{OMbarpsi}), which 
take much simpler forms: 
\begin{equation} \label{fpsipreg}
\begin{split}
\sum\limits_{a=\pm} 
F^{\psi_\pm,reg}_{a \bar{a} x}(u+i \pi,u,\theta-i \tfrac{\pi}{2})&=\delta_{x,+} \,{\cal F}_{\pm}(u,\theta), \\
\sum\limits_{a=\pm} 
F^{\bar{\psi}_\pm,reg}_{a \bar{a} x}(u+i \pi,u,\theta-i \tfrac{\pi}{2})&=\mp \, i\, \delta_{x,-} \,{\cal F}_{\mp}(u,\theta),
\end{split}
\end{equation}
where the weak coupling expansion of ${\cal F}_{\pm}(u,\theta)$ takes the form:
\begin{equation} \label{cF}
\begin{split}
{\cal F}_\pm(u,\theta)=g\,{\cal F}_\pm^{(1)}(u,\theta)+O(g^2)
\end{split}
\end{equation}
with
\begin{equation} \label{cFOg}
\begin{split}
{\cal F}_\pm^{(1)}(u,\theta)=\pm \frac{\left(\frac{1}{4}\pm\frac{i}{4}\right)
   \sqrt{m} e^{\pm \frac{\theta }{2}\mp u} \tanh
   (u-\theta ) \text{sech}(u-\theta )}{\sqrt{2}
   \pi ^{3/2}}.
\end{split}
\end{equation}
When computing the weak coupling expansion of the regularized form-factors, we used the 
perturbative expansion of the S-matrix elements, too. Just for completeness we present 
that of the least nontrivial part of (\ref{Sm}):
 \begin{equation} \label{apert}
\begin{split}
a(\theta)=1-i \,g \,\tanh \left(\frac{\theta }{2}\right)+O(g^2).
\end{split}
\end{equation}
For the perturbative test of our final formulas (\ref{fapsi}), (\ref{fbarapsi}), 
the weak coupling expansion of the function 
$U(\theta)$ is also necessary. It takes the form: 
\begin{equation} \label{Ufvpert}
\begin{split}
U(\theta)=\frac{2 \, g}{\cosh(\theta )}+O(g^2).
\end{split}
\end{equation}

\subsection{Perturbative results from integrability}

In this subsection, with help of our integrability based formulas (\ref{fapsi}), (\ref{fbarapsi}), 
 we compute the leading order term in the weak coupling expansion of 
the L\"uscher-correction to the propagator of the massive Thirring model. 
At this point, one has to take carefully into account, that 
the L\"uscher-correction is a correction to the Bethe-Yang limit, which also has $L$ dependence, 
even though it is only polynomial in the inverse of the volume. 
To use a parameterization comparable to direct field theory computations, 
we rewrite the Bethe-Yang limit of the rapidity (\ref{thBY}) in the following way:
\begin{equation} \label{thBYuj}
\begin{split}
\sinh \theta_{BY}=\tfrac{Q}{\ell}, \qquad \mbox{with} \qquad Q=q\, L\in (2 {\mathbb Z}-1)\,\pi, \quad 
\ell=m L.
\end{split}
\end{equation}
The parameter $Q$ is a dimensionless quantum number, reflecting the anti-periodic boundary condition. 
In principle, it is an odd integer multiple of $\pi,$ but in the sequel we will consider, as if it was 
 an arbitrary complex parameter. With the help of (\ref{Gkifejt1}) and (\ref{dfapsi}), (\ref{dfbarapsi}) 
the L\"uscher-correction to the residue of the fermion propagator (\ref{Gmom}) can be written as follows:
\begin{equation} \label{resGlus}
\begin{split}
\underset{\omega=i \, {\cal E}(q)}{\mbox{Res}} \, \Gamma_{\alpha \beta}^{(L)}(\omega,q)=\!
\frac{i}{\rho_1(\theta_{BY})} {\bigg{\{}} \!-\!\frac{F^{\psi_\beta}_+(\theta_{BY})\,\bar{F}^{\bar{\psi}_\alpha}_+(\theta_{BY})}{2 \pi}\,
\bigg[ \frac{\sinh \theta_{BY}}{\cosh^2 \theta_{BY}} \, \Omega_0(\theta_{BY})\!+\!
\frac{\Omega_1(\theta_{BY})}{\cosh \theta_{BY}} \bigg]+\\
+\frac{\Omega_0(\theta_{BY})}{2 \pi \cosh \theta_{BY}}\, 
\frac{d}{d \theta}\bigg[ F^{\psi_\beta}_+(\theta)\,\bar{F}^{\bar{\psi}_\alpha}_+(\theta) \bigg]\!\Bigg|_{\theta_{BY}} \!\!\!\!\!+ F^{\psi_\beta}_+(\theta_{BY})\, \Omega^{\bar{\psi}_\alpha}_+(\theta_{BY})+
\bar{F}^{\bar{\psi}_\alpha}_+(\theta_{BY}) \, \Omega^{{\psi}_\beta}_+(\theta_{BY}) {\bigg{\}}}.
\end{split}
\end{equation}
Using the weak coupling expansions of the previous subsection, the 1-particle form-factors (\ref{1ffnorm}) and the definitions (\ref{rho1}) and (\ref{OM0})-(\ref{OMbarpsi}), the 1-loop result can be given by a simple formula:
\begin{equation} \label{R1loopint}
\begin{split}
\underset{\omega=i \, {\cal E}(q)}{\mbox{Res}} \, \Gamma_{++}^{(L)}(\omega,q)=\!
\underset{\omega=i \, {\cal E}(q)}{\mbox{Res}} \, \Gamma_{--}^{(L)}(\omega,q)=\!
g\,\frac{i \, m}{\pi} \frac{Q^2}{(Q^2+\ell^2)^{3/2}} \, K_0(\ell)+O(g^2), \\
\underset{\omega=i \, {\cal E}(q)}{\mbox{Res}} \, \Gamma_{-+}^{(L)}(\omega,q)=\!
-\underset{\omega=i \, {\cal E}(q)}{\mbox{Res}} \, \Gamma_{+-}^{(L)}(\omega,q)=\!
g\, \frac{i \, m}{\pi} \frac{Q \, \ell}{(Q^2+\ell^2)^{3/2}} \, K_0(\ell)+O(g^2),
\end{split}
\end{equation}
where $K_0(\ell)$ is the modified Bessel-function.

\section{The Lagrangian perturbation theory} \label{sect7}

Now, we compute the finite volume corrections to the fermion propagator at the 1-particle pole 
from field theoretical perturbation theory upto 1-loop order. 
The starting  point is the Euclidean action (\ref{Seucl}), that we recall, but for regularization reasons 
in $d$-dimensions:
\begin{equation} \label{Sdim}
\begin{split}
S[\bar{\psi},\psi]=\int\limits \! d^d x \, {\big\{} 
-\bar{\psi}\big( \gamma_\mu \partial_\mu+m_0 \big)\psi
+\tfrac{g}{2} \, \bar{\psi}\gamma_\mu \psi \, \bar{\psi}\gamma_\mu \psi
{\big\}}.
\end{split}
\end{equation}
Here $m_0$ denotes the bare mass and now the Euclidean $\gamma$-matrices satisfy the $d$-dimensional 
Clifford-algebra:
\begin{equation} \label{cliffd}
\begin{split}
{\big{\{}} \gamma_\mu,\gamma_\nu {\big{\}}}=2 \, \delta_{\mu \nu}\, {\bf 1}, \qquad \mu,\nu=0,1,..,d-1.
\end{split}
\end{equation}
In our actual computations, we will use $\mbox{Tr} \, {\bf 1}=d,$ for the trace of the unity matrix in the 
spinor space{\footnote{ We note that for $d=2-\epsilon,$ any 
$\mbox{Tr} \, {\bf 1}\!=\!2\!+\!O(\epsilon),$ choice would lead to the same physical results.}}. 
The free propagator at infinite volume takes the usual form \cite{ZJbook}:
\begin{equation} \label{propinf}
\begin{split}
\langle \bar{\psi}_\alpha(x) \psi_\beta(y)\rangle_0=\Delta_{\beta \alpha}(y-x), \qquad 
\Delta(y-x)=\frac{1}{(2\pi)^d} \int \! d^d p \, e^{i \, p (y-x)} \, \frac{m_0 {\bf 1}-i \slashed{p}}{p^2+m^2_0},
\end{split}
\end{equation}
where $\slashed{p}=\gamma_\mu \, p_\mu,$ as usual.
Now, we compute perturbatively the finite volume 2-point function (\ref{Gmom}):
\begin{equation} \label{Gmomuj}
\begin{split}
\Gamma_{\alpha \beta}(\omega,q)=\!\frac{1}{L} \, \int\limits_{-L/2}^{L/2}\!dx_1 \!\!
\int\limits d^{d-2}\! x_{\bot} \!\!\!
\int\limits_{-\infty}^{\infty} \! dt \, e^{i\, \omega t+i \, q {\bf x}} \, 
\langle \bar{\psi}_\alpha({\bf x},t) \psi_{\beta}(0) \rangle_L, \qquad {\bf x}=(x_1,x_{\bot})
\end{split}
\end{equation}
where antiperiodic boundary conditions are imposed on the fermion fields along the compactified 
$x_1$-direction. 

A key ingredient to the perturbative computations is the free propagator corresponding to the 
 boundary conditions under consideration. A simple computation shows, that the $d$-dimensional  
free propagator being  antiperiodic with respect to $L$ in the direction $\mu=1,$ takes the form as follows:   
\begin{equation} \label{DA}
\begin{split}
\Delta_A(x)=\sum\limits_{n \in \mathbb{Z}} (-1)^n \int\limits \!\frac{d^d p}{(2\pi)^d}\, 
e^{i \, p \, x+i \, p_1  n L} \frac{m_0 {\bf 1}-i \slashed{p}}{p^2+m^2_0}.
\end{split}
\end{equation}
This form implies, that in perturbation theory everything is the same as in the usual infinite volume case, 
apart from a change in the "integration measure":
\begin{equation} \label{intint}
\begin{split}
 \int\limits \!\frac{d^d p}{(2\pi)^d}...\, 
 \rightarrow \int\limits_L \!\frac{d^d p}{(2\pi)^d}...=
\sum\limits_{n \in \mathbb{Z}} (-1)^n \int\limits \!\frac{d^d p}{(2\pi)^d}\,
e^{i \, p \, x+i \, p_1  n L}...
\end{split}
\end{equation}
Simple computation shows, that at 1-loop order the propagator in momentum space take the form: 
\begin{equation} \label{proppert}
\begin{split}
\Gamma_{\alpha \beta }(\omega,q)=\frac{1}{L}
\,\bigg(\tilde{\Delta}_0(p)+\tilde{\Delta}_0(p)\,\Gamma^{(1)}(p) \,\tilde{\Delta}_0(p)\bigg)_{\beta \alpha},
\quad \tilde{\Delta}_0(p)=\frac{m_0-i \slashed{p}}{p^2+m_0^2}, \quad p=(\omega,q).
\end{split}
\end{equation}
We note, that as a consequence of antiperiodic boundary conditions the allowed set of values for the 
momentum $q\in \tfrac{2\mathbb{Z}-1}{L} \pi.$
The actually momentum independent $\Gamma^{(1)}(p)$ is of the form:  
\begin{equation} \label{Gf1}
\begin{split}
\Gamma^{(1)}(p)=g\, \bigg(\mbox{Tr}[\gamma_\mu \Delta_A(0)]\, \gamma_\mu-\gamma_\mu \Delta_A(0) \gamma_\mu \bigg).
\end{split}
\end{equation}
Because of the momentum independence of $\Gamma^{(1)}(p),$ in the sequel we will emphasize
 its dependence on the bare parameters $m_0$ and $\epsilon$ instead of the apparent $p$-dependence.  
Using the integral representation (\ref{DA}), and the identity $\mbox{Tr}\, \gamma_\mu=0,$ $\Gamma^{(1)}$ 
take the simple integral representation as follows:
\begin{equation} \label{Gf1uj}
\begin{split}
\Gamma^{(1)}(m_0,\epsilon)=-g\,d\, m\, I_1(m_0,\epsilon)\, {\bf 1},
\end{split}
\end{equation}
where
\begin{equation} \label{I_1}
\begin{split}
 I_1(m_0,\epsilon)=\sum\limits_{n \in \mathbb{Z}} (-1)^n \int\limits \!\frac{d^d p}{(2\pi)^d}\, 
\frac{e^{i \, p \, x+i \, p_1  n L} }{p^2+m^2_0}, \qquad \mbox{with} \quad d=2-\epsilon.
\end{split}
\end{equation}
Evaluating the integrals it takes the form:
\begin{equation} \label{I1QQ}
\begin{split}
I_1(m_0,\epsilon)&=Q_\infty(m_0,\epsilon)+Q_L(m_0,\epsilon), \\
Q_\infty(m_0,\epsilon)&=\frac{1}{2 \pi \epsilon}+\frac{\log(4 \pi)-\log(m^2/\kappa^2)-\gamma_E}{4 \pi}+O(\epsilon), \\
Q_L(m_0,\epsilon)&=-\frac{1}{2 \pi} \sum\limits_{n \in \mathbb{Z}} \, K_0(|n|\, m_0\, L)+O(\epsilon),
\end{split}
\end{equation} 
where $\gamma_E$ is the Euler-gamma, $K_0$ is the modified Bessel-function and $\kappa$ is a mass 
scale emerging as a consequence of the regularization method.
In (\ref{I1QQ}) $Q_\infty$ corresponds to the infinite volume part and $Q_L$ is the finite volume part, which 
upto L\"uscher-order can be written as follows: 
\begin{equation} \label{QL}
\begin{split}
Q_L(m_0,0)=-\frac{1}{\pi} K_0(m_0\, L)+O(e^{-2 m_0 L}).
\end{split}
\end{equation}
It is easy to see, that at 1-loop order, wave-function renormalization is not required, but mass renormalization 
is necessary to define the renormalized propagator. The simple steps of the 1-loop mass renormalization can be 
in the easiest way done at the inverse of the propagator:  
\begin{equation} \label{Dinv}
\begin{split}
\Gamma_{\alpha \beta}^{-1}(p)=L \, \bigg( m_0 \, {\bf 1}+i \, \slashed{p}-\Gamma^{(1)}(m_0,\epsilon) \, {\bf 1}+O(g^2) 
\bigg)_{\beta \alpha},
\end{split}
\end{equation}
by inserting $m_0=m+\delta m,$ with $m$ and $\delta m$ being the physical mass and the additive 
mass renormalization constant, respectively. The latter is determined to cancel the divergences from 
(\ref{Dinv}), such that at infinite volume, the location of the pole of the propagator is at $p^2=-m^2.$ 
Thus, upto $O(g)$ the renormalized propagator inverse will take the form:  
\begin{equation} \label{DR}
\begin{split}
\Gamma_{\alpha \beta}^{-1}(p)&=m \, {\bf 1}+i \, \slashed{p}-\Gamma^{(1)}_L(m) \, {\bf 1}+O(g^2), \\
\Gamma^{(1)}_L(m)&=2\,g\,m\, Q_L(m,0)=-\frac{2\, g\, m}{\pi}\,K_0(\ell)+O(e^{-2 \ell}), 
\qquad \ell=m\, L,
\end{split}
\end{equation}
and the mass renormalization constant:
\begin{equation} \label{dm}
\begin{split}
\delta m=-g\, (2-\epsilon)\, m\,Q_\infty(m,\epsilon)+O(g^2). 
\end{split}
\end{equation}
From the position of the pole of the renormalized propagator (\ref{DR}), the 1-loop finite volume shift of the 1-particle energy denoted by $\delta {\cal E}(q),$ can be determined:   
\begin{equation} \label{polecond}
\begin{split}
(m-\Gamma^{(1)}_L(m))^2+\sum_\mu p_\mu \, p_\mu+O(g^2)=0,
\end{split}
\end{equation}
where
\begin{equation} \label{p_mu}
\begin{split}
p_\mu=\begin{pmatrix}  i({\cal E}(q)+\delta {\cal E}(q))\\ q
\end{pmatrix}, \qquad {\cal E}(q)=\sqrt{q^2+m^2}.
\end{split}
\end{equation}
Finally, one gets:
\begin{equation} \label{dEq}
\begin{split}
\delta {\cal E}(q)=-\frac{m}{{\cal E}(q)} \, \Gamma^{(1)}_L(m)+O(g^2).
\end{split}
\end{equation}
To extract the form-factors, the residue at the finite volume 1-particle pole should be taken 
(see (\ref{1partGMv1})). This can be done, by taking the inverse of (\ref{DR}) at
\begin{equation} \label{p_muuj}
\begin{split}
p_\mu=\begin{pmatrix}  i({\cal E}(q)+\delta {\cal E}(q))+\Delta E\\ q
\end{pmatrix}, \qquad \Delta E=\omega-i({\cal E}(q)+\delta {\cal E}(q)),
\end{split}
\end{equation} and extract the term proportional to $\tfrac{1}{\Delta E}.$ After a simple 
computation one ends up with the following result for the renormalized propagator:
\begin{equation} \label{DRDE}
\begin{split}
\Gamma_{\alpha \beta}(\omega,q)=
\frac{1}{L\, \Delta E}\bigg(\frac{m-i \slashed{p}^{(\infty)}}{2 \, i \, {\cal E}(q)}+
R_{\bf 1}^{(L)}{\bf 1}+\sum_{\mu=0,1} \gamma_\mu \, R_{\gamma_\mu}^{(L)}+O(g^2)\bigg)_{\beta \alpha}+...,
\end{split}
\end{equation}
where
\begin{equation} \label{R1L}
\begin{split}
R_{\bf 1}^{(L)}=-\frac{m \, \delta {\cal E}(q)}{2 \, i\, {\cal E}^2(q)},  \qquad
 R_{\gamma_\mu}^{(L)}=-\frac{\delta {\cal E}(q)}{2 \, {\cal E}(q)}
\bigg(i\, \delta_{\mu 0}\, - \frac{p_\mu^{(\infty)}}{ {\cal E}(q)} \bigg),
\qquad \quad
p_\mu^{(\infty)}=\begin{pmatrix}  i\,{\cal E}(q) \\ q
\end{pmatrix},
\end{split}
\end{equation}
and the dots in (\ref{DRDE}) stand for terms $O(1)$ in $\Delta E.$ The term being proportional to 
$m-i \slashed{p}^{(\infty)}$ describes the infinite volume part of the form-factors, while  the rest 
corresponds to the finite volume corrections. With the help of (\ref{DR}) and (\ref{dEq}), the finite 
volume corrections can be expressed in terms of the $K_0(\ell)$ Bessel-function. The main ingredients 
take the form: 
\begin{equation} \label{R1L}
\begin{split}
\delta{\cal E}(q)=-g\, \frac{2\, m^2 \, K_0(\ell)}{{\pi\, \cal E}(q)}, \quad
R_{\bf 1}^{(L)}=g \frac{i}{\pi} \frac{Q^2\, \ell \, K_0(\ell)}{(Q^2+\ell^2)^{3/2}},  \qquad
R_{\gamma_\mu}^{(L)}=-g \, \delta_{\mu 1}\frac{1}{\pi} \frac{Q\, \ell^2 \, K_0(\ell)}{(Q^2+\ell^2)^{3/2}},
\end{split}
\end{equation}
where again, we used the variables $Q=q\, L$ and $\ell=m\, L.$
Inserting (\ref{R1L}) into (\ref{DRDE}), the $O(g)$ L\"uscher-correction to the residue of the fermion 
propagator will take the form{{\footnote{Here $\Gamma_{\alpha \beta}^{(L)}$ denotes the finite volume correction 
to the infinite volume propagator.}}}:
\begin{equation} \label{Resprop}
\begin{split}
\underset{\omega=i \Delta E_1(L)}{\mbox{Res}}\, \Gamma_{+ +}^{(L)}(\omega,q)&=
\underset{\omega=i \Delta E_1(L)}{\mbox{Res}}\, \Gamma_{- -}^{(L)}(\omega,q)=
g \frac{i}{\pi L} \frac{Q^2\, \ell \, K_0(\ell)}{(Q^2+\ell^2)^{3/2}}+O(g^2,e^{-2\ell}), \\
\underset{\omega=i \Delta E_1(L)}{\mbox{Res}}\, \Gamma_{- +}^{(L)}(\omega,q)&=
-\underset{\omega=i \Delta E_1(L)}{\mbox{Res}}\, \Gamma_{+ -}^{(L)}(\omega,q)=
g \frac{i}{\pi L} \frac{Q\, \ell^2 \, K_0(\ell)}{(Q^2+\ell^2)^{3/2}}+O(g^2,e^{-2 \ell}),
\end{split}
\end{equation}
which agrees with the formula coming from the 1-particle form-factor L\"uscher-formula (\ref{R1loopint}).

\section{Summary and conclusions} \label{summary}

In this paper, using the field theoretical approach of \cite{BBCL}, we derived the leading exponentially 
small in volume corrections to the 1-particle form-factors of a non-diagonally scattering 
relativistic integrable quantum field theory. Our final formulas were tested  
against Lagrangian perturbation theory at 1-loop order in the massive Thirring-model, and 
perfect agreement was found. Our results can be considered as extensions of those of \cite{BBCL} to 
more general cases. In \cite{BBCL}, 1-particle form-factors of operators with zero Lorentz-spin in a bosonic 
diagonally scattering relativistic integrable quantum field theory were considered. Our final results 
also valid for operators with nonzero Lorentz spin, and for both bosonic and fermionic non-diagonally 
scattering relativistic integrable quantum field theories. 

The main result of the paper is the formula (\ref{fapsi}) with the definitions (\ref{fBYjel}) and 
(\ref{OM0})-(\ref{OMbarpsi}), giving the finite volume 1-particle form-factor 
upto leading exponentially small in volume corrections. The physical interpretation of the formula is 
the same as it is in the diagonally scattering case \cite{BBCL}. It is composed of two terms. The 1st one, is 
similar to the formula being valid in the Bethe-Yang limit, but the exponentially small in volume corrections 
to the particle's rapidity and to the density of states are taken into account. The 2nd term is a new type 
of term. It contains a 3-particle form-factor corresponding to the contribution of a virtual particle traveling 
around the world. 

Though, our results give the L\"uscher-corrections only to 1-particle form-factors, we hope, that similarly to 
the case of diagonally scattering theories \cite{BBCL1}, these results could be extended to arbitrary matrix elements 
of local operators in non-diagonally scattering integrable quantum field theories.

\subsection*{Acknowledgements}

The author would like to thank Zolt\'an Bajnok and J\'anos Balog for useful discussions, and \'Ad\'am 
Andr\'as Kelemen for participating at the early stage of this work.
This work was supported by the NKFIH research Grant K134946.

\appendix

\section{List of $H$ and $G$ functions} \label{appA}

Here we list the tensor functions entering (\ref{PIuj}) and (\ref{J0}). 
The $H$ functions are building blocks of the regularized matrix element 
$\Pi(u|\beta_1,\beta_2)_{a b_1 b_2}[F]$ and so they are functionals of the $F$ form factors.
\begin{equation} \label{Hs}
\begin{split}
H_0(u|\beta_1,\beta_2)_{a b_1 b_2}&=s_0 \, F^c_{\bar{a} b_1 b_2}(u+i \pi,\beta_1,\beta_2), \\
H_1^{(-)}(\beta_1,\beta_2)_{a b_1 b_2}&=-s_0 \, \frac{i}{2 \pi} \, \sum\limits_{x} 
S_{b_1 b_2}^{a x}(\beta_1-\beta_2) \, F_{x}(\beta_2), \\
H_1^{(+)}(\beta_1,\beta_2)_{a b_1 b_2}&=s_0 \, \frac{i}{2 \pi} \, \delta_{a b_1} \, F_{b_2}(\beta_2), \\
H_2^{(-)}(\beta_1,\beta_2)_{a b_1 b_2}&=-\frac{i}{2 \pi} \, \delta_{a b_2} \, F_{b_1}(\beta_1), \\
H_2^{(+)}(\beta_1,\beta_2)_{a b_1 b_2}&= \frac{i}{2 \pi} \, \sum\limits_{x} 
S_{b_1 b_2}^{x a}(\beta_1-\beta_2) \, F_{x}(\beta_1),
\end{split}
\end{equation}
The $G$ functions are building blocks of the regularized matrix element 
$\bar{\Pi}(u|\beta_1,\beta_2)_{a b_1 b_2}[f]$ and so they are functionals of the $f$ form factors;
\begin{equation} \label{Gs}
\begin{split}
G_0(u|\beta_1,\beta_2)_{b_2 b_1 a}&=s_0 \, s_f \, f^c_{{a} \bar{b}_2 \bar{b}_1}(u+i \pi,\beta_2,\beta_1), \\
G_1^{(-)}(\beta_1,\beta_2)_{b_2 b_1 a}&=- \frac{i}{2 \pi} \,s_f \, \delta_{a b_1} \, f_{\bar{b}_2}(\beta_2), \\
G_1^{(+)}(\beta_1,\beta_2)_{b_2 b_1 a}&= \frac{i}{2 \pi} \,s_f \, \sum\limits_{x} 
S_{b_2 b_1}^{\bar{x} a}(\beta_2-\beta_1) \, f_{x}(\beta_2), \\
G_2^{(-)}(\beta_1,\beta_2)_{b_2 b_1 a}&= -\frac{i}{2 \pi} s_0 \, s_f \, \sum\limits_{x} 
S_{b_2 b_1}^{a x}(\beta_2-\beta_1) \, f_{\bar{x}}(\beta_1), \\
G_2^{(+)}(\beta_1,\beta_2)_{b_2 b_1 a}&=\frac{i}{2 \pi} \,s_0 \, s_f \, \delta_{a b_2} \, f_{\bar{b}_1}(\beta_1).
\end{split}
\end{equation}


\begin{thebibliography}{10}
\expandafter\ifx\csname url\endcsname\relax
  \def\url#1{\texttt{#1}}\fi
\expandafter\ifx\csname urlprefix\endcsname\relax\def\urlprefix{URL }\fi
\expandafter\ifx\csname href\endcsname\relax
  \def\href#1#2{#2} \def\path#1{#1}\fi


\bibitem{Beisert:2010jr}
  N.~Beisert {\it et al.},
  {\it Review of AdS/CFT Integrability: An Overview,}
  Lett.\ Math.\ Phys.\  {\bf 99}, 3 (2012)

%

\bibitem{KoEss}
  F.~H.~L.~Essler and R.~M.~Konik,
  ``Applications of massive integrable quantum field theories to problems in condensed matter physics,''
  In *Shifman, M. (ed.) et al.: From fields to strings, vol. 1* 684-830,
  [cond-mat/0412421 [cond-mat.str-el]].

%

\bibitem{Luscher1}
M.~L\"uscher,
{\it Volume Dependence of the Energy Spectrum in Massive Quantum Field
Theories. 1. Stable Particle States,}
Commun.\ Math.\ Phys.\  {\bf 104} (1986) 177.

\bibitem{Luscher2}
  M.~L\"uscher,
{\it Volume Dependence of the Energy Spectrum in Massive Quantum Field Theories. 2. Scattering States,}
  Commun.\ Math.\ Phys.\  {\bf 105}, 153 (1986).

%

\bibitem{KlMe}
T.~R.~Klassen and E.~Melzer,
{\it On the relation between scattering amplitudes and finite size mass
corrections in QFT,}
Nucl.\ Phys.\ B {\bf 362} (1991) 329.

\bibitem{JaLu}
R.~A.~Janik and T.~Lukowski,
{\it Wrapping interactions at strong coupling: The Giant magnon,}
Phys.\ Rev.\ D {\bf 76} (2007) 126008
[arXiv:0708.2208 [hep-th]].

%

\bibitem{BaJa}
Z.~Bajnok and R.~A.~Janik,
{\it Four-loop perturbative Konishi from strings and finite size effects
for multiparticle states,}
Nucl.\ Phys.\ B {\bf 807} (2009) 625

\bibitem{Bombardelli:2013yka}
  D.~Bombardelli,
  {\it A next-to-leading Luescher formula,}
  JHEP {\bf 1401}, 037 (2014)


\bibitem{Zamolodchikov:1989cf}
  A.~B.~Zamolodchikov,
 {\it Thermodynamic Bethe Ansatz in Relativistic Models. Scaling Three State Potts and Lee-yang Models,}
  Nucl.\ Phys.\ B {\bf 342}, 695 (1990).

\bibitem{Dorey:1996re}
  P.~Dorey and R.~Tateo,
  {\it Excited states by analytic continuation of TBA equations,}
  Nucl.\ Phys.\ B {\bf 482}, 639 (1996)

%

\bibitem{BHO34}
J. Balog, \'A. Heged\H{u}s,
{\it TBA equations for excited states in the $O(3)$ and $O(4)$ nonlinear $\sigma$-models,} 
 J. Phys.{\bf  A37} (2004) 1881-1901.

%

\bibitem{GKVsu2}
  N.~Gromov, V.~Kazakov and P.~Vieira,
  ``Finite Volume Spectrum of 2D Field Theories from Hirota Dynamics,''
  JHEP {\bf 0912} (2009) 060
  [arXiv:0812.5091 [hep-th]].

%

\bibitem{KP1}
A. Kl\"umper,M. Batchelor, P. Pearce,
``Central charges of the 6- and 19-vertex models with twisted boundary conditions,''
{\em J. Phys. A.}{\bf 24} (1991) 3111.

\bibitem{ddv92}C. Destri and H.J. de Vega, 
``New thermodynamic Bethe ansatz equations without strings,'' 
{\emph Phys.Rev.Lett.}{\bf 69 }(1992) 2313-2317.
[hep-th/9203064].

%

\bibitem{SuzS1}
J. Suzuki,
{\it "Excited states nonlinear integral equations for an integrable anisotropic spin 1 chain"},
J.Phys.{\bf  A37} (2004) 11957-11970,
[arXiv:hep-th/0410243[hep-th]]

%
  \bibitem{KLsun}
  V.~Kazakov and S.~Leurent,
  {\it ``Finite size spectrum of $SU(N)$ principal chiral field from discrete Hirota dynamics,''}
  Nucl.\ Phys.\ B {\bf 902} (2016) 354
  [arXiv:1007.1770 [hep-th]].

%
\bibitem{BHnlie}
J. Balog, \'A. Heged\H{u}s, {\it {"Hybrid-NLIE for the AdS/CFT spectral problem"}}
{\em JHEP} {\bf 1208} (2012) 022.

%
\bibitem{QSC54}
N. Gromov, V. Kazakov, S. Leurent, D. Volin,
{\it "Quantum spectral curve for arbitrary state/operator in $AdS_5/CFT_4$"},
JHEP {\bf 1509}, 187 (2015),
[arXiv:  1405.4857[hep-th]]
%
\bibitem{QSC43}
D. Bombardelli, A. Cavagli\'a, D. Fioravanti, N. Gromov, R. Tateo,
{\it "The full Quantum Spectral Curve for $AdS_4/CFT_3$" },
JHEP {\bf 1709}, 140 (2017),
[arXiv: 1701.00473 [hep-th]]
%
\bibitem{PT10T}
  B.~Pozsgay and G.~Tak\'acs,
 {\it ``Form factor expansion for thermal correlators,'' }
 {{\em J.\ Stat.\ Mech.\ }   {\bf 1011} (2010) P11012},
  [arXiv:1008.3810 [hep-th]].
%
\bibitem{Gohmann}
C.~Babenko, F.~G\"ohmann, K. K.~Kozlowski and J. Suzuki, 
{\it ``A thermal form factor series for the longitudinal two-point function of the 
Heisenberg-Ising chain in the antiferromagnetic massive regime,'' }
 {{\em J.\ Math.\ Phys.\ }   {\bf 62} (2021) 4, 041901},
  [arXiv:2011.12752 [cond-mat,stat-mech]].

%
\bibitem{BJsftv}
  Z.~Bajnok and R.~A.~Janik,
{\it  ``String field theory vertex from integrability,'' }
  {\em JHEP  1504} (2015) 042,
  [arXiv:1501.04533 [hep-th]].
%
\bibitem{KomatsuLect}
  S.~Komatsu,
  {\it Lectures on Three-point Functions in N=4 Supersymmetric Yang-Mills Theory,}
  arXiv:1710.03853 [hep-th].

%
\bibitem{PT08a}
B.~Pozsgay and G.~Tak\'acs, ``{Form-factors in finite volume I: Form-factor
  bootstrap and truncated conformal space},''
{{\em Nucl.Phys.} {\bf B788} (2008) 167--208}, [arXiv:0706.1445 [hep-th]].
%
\bibitem{PT08b}
B.~Pozsgay and G.~Tak\'acs, ``{Form factors in finite volume. II. Disconnected
  terms and finite temperature correlators},''
{{\em Nucl.Phys.} {\bfseries B788} (2008) 209--251},
 [arXiv:0706.3605 [hep-th]].

%

\bibitem{LM99}
A.~Leclair and G.~Mussardo, ``{Finite temperature correlation functions in
  integrable QFT},''
  {\em Nucl.Phys.}{\bf B552} (1999) 624--642, 
[arXiv:9902075 [hep-th]].

%

\bibitem{PST14}
B. Pozsgay, I.~Sz\'ecs\'enyi and G.~Tak\'acs, ``{Exact finite volume expectation values of local operators 
in excited states},''
{\em JHEP 1504} (2015) 023,
[arXiv:1412.8436 [hep-th]].  

%
\bibitem{PSZLM}
B. PozsgayI. M. Sz\'ecs\'enyi, 
{\it "LeClair-Mussardo series for two-point functions in Integrable QFT"},
JHEP {\bf 1805}, 170 (2018),
[arXiv:1802.05890 [hep-th]]

%

\bibitem{SmirNeg}
  S.~Negro and F.~Smirnov,
  ``On one-point functions for sinh-Gordon model at finite temperature,''
  Nucl.\ Phys.\ B {\bf 875} (2013) 166,
  [arXiv:1306.1476 [hep-th]].
%
\bibitem{SmirBajn}
  Z.~Bajnok and F.~Smirnov,
  ``Diagonal finite volume matrix elements in the sinh-Gordon model,''
  Nucl.\ Phys.\ B {\bf } (2019) 114664,
  [arXiv:1903.06990 [hep-th]].

%
\bibitem{Pomu}
  B.~Pozsgay,
  Nucl.\ Phys.\ B {\bf 802} (2008) 435
  doi:10.1016/j.nuclphysb.2008.04.021
%
\bibitem{BBCL}
Z.~Bajnok, J.~Balog, M.~L\'ajer and C.~Wu,
  ``Field theoretical derivation of L\"uscher's formula and calculation of finite volume form factors,''
  JHEP {\bf 1807} (2018) 174, 
  [arXiv:1802.04021 [hep-th]].
  
 \bibitem{BBCL1}
  Z.~Bajnok, M.~L\'ajer, B.~Sz\'epfalvi and I.~Vona,
  ``Leading exponential finite size corrections for non-diagonal form factors,''
  JHEP {\bf 1907} (2019) 173
  doi:10.1007/JHEP07(2019)173
  [arXiv:1904.00492 [hep-th]]. 

%

\bibitem{SGV}
  M.~Jimbo, T.~Miwa and F.~Smirnov,
  ``Hidden Grassmann structure in the XXZ model V: Sine-Gordon model,''
  Lett.\ Math.\ Phys.\  {\bf 96} (2011) 325,
  [arXiv:1007.0556 [hep-th]].  
%

\bibitem{SGen}
\'A. Heged\H{u}s, 
{\it "Finite volume expectation values in the sine-Gordon model" },
Nucl. Phys. {\bf B948} 114749
[arXiv:1901.01806[hep-th]]

%

\bibitem{SmirBab}
  C.~Babenko and F.~Smirnov,
  ``One point functions of fermionic operators in the Super Sine Gordon model,''
  arXiv:1905.09602 [hep-th].  

%

%
\bibitem{HuPoPri}
A. Hutsalyuk, B. Pozsgay, L. Pristy\'ak,
{\it "The LeClair-Mussardo series and nested Bethe Ansatz"},
Nucl.Phys. {\bf B 964} (2021) 115306,
[arXiv: 2009.13203 [cond-mat]]

%

\bibitem{Smirnov}
  F.~A.~Smirnov,
  {\it Form-factors in completely integrable models of quantum field theory,}
  Adv.\ Ser.\ Math.\ Phys.\  {\bf 14}, 1 (1992).

%



%

\bibitem{LUK1}
 S. L. Lukyanov, ``{Free field representation for massive
integrable models},''
{{\em Commun. Math. Phys.}
{\bf 167} (1995) 183--226},
[arXiv:hep-th/9307196].

%

\bibitem{LUK2}
 S. L. Lukyanov, ``{Form factors of exponential fields in
the sine-Gordon model},''
{{\em Mod. Phys. Lett.} {\bf A12} 
(1997) 2543--2550},
[arXiv:hep-th/9703190].

%

\bibitem{BABK1}
 H. M. Babujian, A. Fring, M. Karowski, and
A. Zapletal, ``{Exact form factors in integrable quantum
field theories: The sine-Gordon model},''
{{\em Nucl. Phys.} {\bf B538} 
(1999) 535--586},
[arXiv:hep-th/9805185].

%

\bibitem{BABK2}
 H. Babujian and M. Karowski, ``{Exact form factors in
integrable quantum field theories: The sine-Gordon
model. II},''
{{\em Nucl. Phys.}  {\bf B620} 
(2002) 407--455},
[arXiv:hep-th/0105178].

%

%
\bibitem{ZamiZami}
  A.~B.~Zamolodchikov and A.~B.~Zamolodchikov,
  ``Factorized s Matrices in Two-Dimensions as the Exact Solutions of Certain Relativistic Quantum Field Models,''
  Annals Phys.\  {\bf 120} (1979) 253.

%

\bibitem{FevPhd} G. Feverati, ``Finite volume spectrum of sine-Gordon model and its restrictions (Phd Thesis),''
[hep-th/0001172].

%

\bibitem{ZJbook} P. Zinn Justin, ``Quantum field theory and critical phenomena''
{\em Oxford, 1981,  ISBN-13: 978-0198509233 }



\end{thebibliography}
\end{document}